\documentclass[conference]{IEEEtran}

\IEEEoverridecommandlockouts

\usepackage{cite}
\usepackage{amsmath,amssymb,amsfonts}
\usepackage{algorithmic}
\usepackage{graphicx}
\usepackage{textcomp}
\usepackage{xcolor}

\usepackage{mathtools,amsmath}
\usepackage{amsmath,amssymb,amsfonts}
\usepackage{graphicx} 
\usepackage{amsmath}  
\usepackage{graphicx} 
\usepackage{textcmds} 
\usepackage{tikz}
\usepackage{multirow}
\usepackage{subcaption}
\usepackage{array, makecell}%
\newcommand*\circled[1]{\tikz[baseline=(char.base)]{
            \node[shape=circle,fill,inner sep=2pt] (char) {\textcolor{white}{#1}};}}

\def\BibTeX{{\rm B\kern-.05em{\sc i\kern-.025em b}\kern-.08em
    T\kern-.1667em\lower.7ex\hbox{E}\kern-.125emX}}

\title{Towards Efficient LUT-based PIM: A Scalable and Low-Power Approach for Modern Workloads}

\author{\IEEEauthorblockN{Bahareh Khabbazan, Marc Riera, Antonio González }
\IEEEauthorblockA{\textit{dept. of Computer Architecture } \\
\textit{Universitat Polit\`{e}cnica de Catalunya (UPC)}\\
Barcelona, Spain\\
\{bahareh.khabbazan, marc.riera.villanueva, antonio.gonzalez\}@upc.edu}
}

\begin{document}

\maketitle

\begin{abstract}
Data movement in memory-intensive workloads, such as deep learning, incurs energy costs that are over three orders of magnitude higher than the cost of computation. Since these workloads involve frequent data transfers between memory and processing units, addressing data movement overheads is crucial for improving performance. Processing-using-memory (PuM) offers an effective solution by enabling in-memory computation, thereby minimizing data transfers. In this paper we propose \textbf{Lama}, a LUT-based PuM architecture designed to efficiently execute SIMD operations by supporting independent column accesses within each mat of a DRAM subarray. Lama exploits DRAM's mat-level parallelism and open-page policy to significantly reduce the number of energy-intensive memory activation (ACT) commands, which are the primary source of overhead in most PuM architectures. Unlike prior PuM solutions, Lama supports up to 8-bit operand precision without decomposing computations, while incurring only a 2.47\% area overhead. Our evaluation shows Lama achieves an average performance improvement of 8.5$\times$ over state-of-the-art PuM architectures and a 3.8$\times$ improvement over CPU, along with energy efficiency gains of 6.9$\times/$8$\times$, respectively, for bulk 8-bit multiplication.

We also introduce \textbf{LamaAccel}, an HBM-based PuM accelerator that utilizes Lama to accelerate the inference of attention-based models. LamaAccel employs exponential quantization to optimize product/accumulation in dot-product operations, transforming them into simpler tasks like addition and counting. LamaAccel delivers up to 9.3$\times/$19.2$\times$ reduction in energy and 4.8$\times/$9.8$\times$ speedup over TPU/GPU, along with up to 5.8$\times$ energy reduction and 2.1$\times$ speedup over a state-of-the-art PuM baseline.

\end{abstract}

\section{Introduction}
Modern workloads, such as big data analytics, deep learning, and graph processing, are characterized by high data parallelism and low computational intensity~\cite{PnM_fpga}. In conventional systems, like GPUs and CPUs, these applications require frequent data transfers between memory and the processing units, significantly degrading both energy efficiency and performance. The cost of moving data can substantially outweigh the cost of performing computations. This imbalance has been highlighted in previous research~\cite{eie}, which reveals that retrieving a 32-bit word from off-chip DRAM incurs an energy cost of 6400 times higher than performing a simple ADD operation in 45 nm technology. As these applications span various domains and become more prevalent, addressing the inefficiencies and performance limitations caused by data movement is crucial for enhancing the overall system efficiency.

Processing-in-Memory (PIM) aims to alleviate the data transfer bottleneck by integrating computational capabilities directly within the memory units~\cite{ghoseprocessing} itself, thereby reducing data movement and improving overall performance~\cite{mutluprocessing}. PIM-based accelerators are classified into two main approaches: 1)~\textbf{Processing-near-Memory (PnM)} where computations take place in processing elements near the memory arrays. The processing elements can be in the same die as the memory array~\cite{pimornot,newton} or in the logic layer of 3D-stacked memory~\cite{DNA-TEQ, qeihan, neurocube}. Although PnM solutions are limited by memory bandwidth, they are well-suited for offloading complex computations. 2)~\textbf{Processing-using-Memory (PuM)} embeds computational capabilities directly within memory cells. This technique allows data to remain in the memory cells while computations are performed in place, eliminating the need for data movement out of memory. However, PuM techniques lack flexibility to support complex operations due to the intrinsic limitations of the memory. Despite the challenges in supporting a wide range of operations, PuM offers significant throughput due to the large parallelism inherent in the memory. While PnM and PuM each offer distinct advantages, a hybrid solution that leverages the strengths of both approaches can optimize performance by balancing the computational complexity with memory bandwidth. This makes it a compelling choice for a wide range of modern workloads.

Numerous PIM designs have been implemented across a diverse array of memory technologies, including static random access memory (SRAM)~\cite{neuralcache}, dynamic random access memory (DRAM)~\cite{drisa, lacc}, and emerging non-volatile memory technologies such as Resistive RAM (ReRAM)~\cite{ReDY}. PuM in the context of commodity DRAM has drawn more attention compared to SRAM. This increased focus is due to the DRAM's larger memory capacity, and the critical need to reduce costly off-chip data transfers to the host. In addition, compared to NVM, DRAM avoids the drawbacks of limited lifespan~\cite{renew} and the high energy costs associated with memory cell writes. PuM in DRAM platforms typically leverages either \textbf{charge-sharing-based}~\cite{ambit} or \textbf{lookup table (LUT)-based}~\cite{pluto} techniques to perform logic and arithmetic operations inside the memory chip.

\begin{figure}[t!]
\centering
\includegraphics[width=0.8\columnwidth]{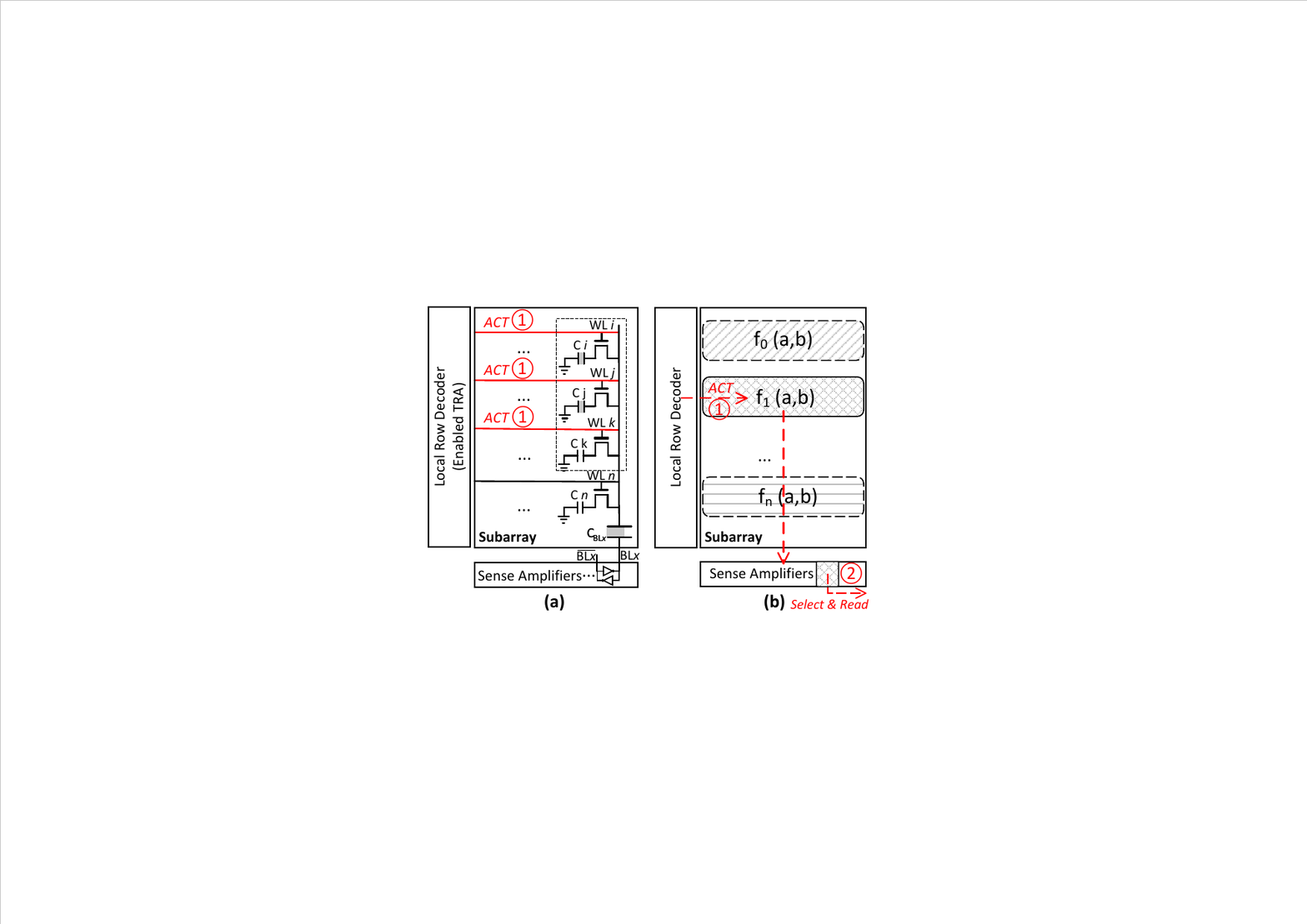}
\caption{Different PuM techniques: (a) charge-sharing (b) LUT-based approach.}
\label{fig:charge_shared_vs_lut}
\end{figure}

Charge-sharing-based techniques, as illustrated in Figure~\ref{fig:charge_shared_vs_lut}a, draw inspiration from Ambit’s Triple Row Activation (TRA) mechanism~\cite{ambit}, offering high compute throughput for bulk bitwise operations. While this approach is effective for simple bitwise tasks, it struggles to efficiently execute more complex arithmetic operations such as multiplication and addition. This inefficiency arises because each arithmetic operation must be decomposed using multiple bitwise operations, requiring numerous memory subarray accesses and costing an extensive amount of cycles and significant energy consumption~\cite{fulcrum}. Additionally, many of these techniques are not fully compatible with commodity DRAM structures, as they demand substantial modifications to the DRAM subarrays~\cite{drisa, dracc} or do not comply with the standard memory command timings~\cite{computedram}.

On the other hand, LUT-based techniques offer a powerful approach for efficiently retrieving results of complex arithmetic operations by leveraging pre-stored data. As depicted in Figure~\ref{fig:charge_shared_vs_lut}b, a LUT essentially functions as a large pre-stored array where each entry corresponds to the result of a specific operation. The operands, often concatenated into a single input value, are used as a unique index or address to quickly access the corresponding result in the LUT. This method provides a more energy-efficient and lower-latency alternative to traditional charge-sharing-based schemes for executing arithmetic operations.

However, simultaneous access to multiple LUT entries, while increasing throughput, can significantly raise energy consumption. This is especially critical in DRAM systems, where energy efficiency is paramount. To effectively enable parallel LUT accesses, it is essential to manage the interleaving of LUT data across different levels of the memory hierarchy. Inefficient mapping of LUT data can lead to sub-optimal access patterns, reducing the system's ability to fully exploit parallel processing. For instance, accessing non-contiguous memory locations across different hierarchy levels (e.g., subarrays, banks) can hinder effective parallelism. To mitigate these challenges, it is crucial to optimize the data layout and employ smart interleaving placement strategies. By strategically distributing LUT entries across the memory hierarchy, one can minimize contention and maximize parallel access capabilities. This ensures that the system can handle multiple LUT queries in parallel, improving overall performance while maintaining energy efficiency.

Several previous works~\cite{pluto, red-LUT, ppim, reconfigurable_pim, lacc, sal-pim} have proposed the design of efficient LUT-based PuM architectures that supports complex bulk arithmetic operations in memory. However, some critical limitations still hinder their practical applicability. The main challenges are:

\textbf{\textit{Area Overhead}}: Executing bulk operations often requires multiple LUT accesses. To achieve this, previous approaches replicate LUTs across different subarrays within a DRAM bank, known as subarray-level parallelism~\cite{salp}. Implementing subarray-level parallelism requires significant modifications to the DRAM bank architecture. Typically, current memory designs allow only one activation (ACT) command at a time. To address this, many proposals replicate row decoders for each subarray, enabling simultaneous ACT commands to different rows within the same bank~\cite{red-LUT, sal-pim}. Additionally, they may introduce extra buffers in some subarrays with sizes equivalent to the row buffer (e.g., 1kB) to store computational results during bulk execution~\cite{pluto}. While these modifications support parallelism, they incur substantial area overheads ranging from 10.2\% to 23.1\%~\cite{pluto}. This added overhead consumes valuable silicon space that could otherwise be allocated for data storage, thereby reducing the memory capacity.


\textbf{\textit{Performance and Energy Inefficiency}}: To handle bulk arithmetic operations, existing LUT-based techniques~\cite{pluto,red-LUT} rely on successive ACT commands across different rows in a DRAM bank, performing a search to retrieve the correct results of the operations. These ACT commands are highly energy-intensive. As an example, in HBM2, the energy required to access a row constitutes 80\% of the total energy consumed for reading one memory block (8 bytes)~\cite{fine-grained}. In addition, these methods are constrained by the Four Activate Window (tFAW) timing parameter of DRAM. This parameter restricts the number of successive ACT commands that can be issued within a given time frame due to DRAM power budget constraints, thereby limiting the parallelism and overall performance of PuM techniques.


\textbf{\textit{Handling Various Arithmetic Precisions}}: Limited support for high-precision arithmetic operations poses significant challenges in LUT-based schemes, especially when handling input operands that exceed 4-bit numerical precision. Many existing LUT-based techniques support only up to 4-bit operand pairs or single 8-bit operands~\cite{reconfigurable_pim}. For higher operand precision, some schemes~\cite{ppim, pluto, lacc} decompose high-precision computations into smaller, lower-precision segments, which can be processed independently across different subarrays. However, this approach adds considerable complexity due to the need for frequent inter-subarray data movement. For example, performing an 8-bit multiplication requires splitting it into four 4-bit multiplications, followed by an 8-stage accumulation process~\cite{ppim}. This accumulation process demands multiple ACT commands and internal data transfers, leading to increased latency and energy consumption. Thus, there is a lack of support for efficient low-cost high-precision arithmetic with PuM techniques.

\textbf{\textit{Exploiting the DRAM hierarchy}}: Each DRAM bank is composed of multiple subarrays and these are further divided into smaller units known as mats~\cite{processingDRAMparadigm}. Mats are constrained by the DRAM circuit design, which requires the mats to transfer consecutive column positions (e.g., 8-bit chunks) via the column select logic (CSL) to the global row buffer~\cite{fulcrum}. Many previous approaches~\cite{red-LUT,reconfigurable_pim, pluto} do not take into account the internal mat structure when deciding their data placement and layout, resulting in reduced parallelism and inefficiencies during LUT accesses.

\textbf{Our goal} in this paper is to overcome the limitations of existing PuM schemes by introducing \textit{Lama}, a \underline{L}ightweight ~\underline{a}daptive~\underline{m}emory~\underline{a}ccess mechanism for executing bulk complex arithmetic operations using a LUT-based PuM approach. To demonstrate Lama's applicability for data-intensive applications, we also propose~\textit{LamaAccel}, an HBM-based PIM accelerator based on Lama to enhance the efficiency of Large Language Models (LLMs) for Machine Learning (ML). Lama redefines current LUT-based PuM designs by enhancing energy efficiency and scalability while maintaining high performance, without requiring changes to DRAM timing parameters or incurring significant area overheads.

Lama introduces an advanced PuM LUT mechanism optimized for arithmetic operations with two operands. To reduce the overhead of successive ACT commands during bulk operations, Lama leverages two key features of commodity DRAM: \textbf{mat-level parallelism} and the \textbf{open-page policy}.

\begin{figure}[t!]
\centering
\includegraphics[width=0.75\columnwidth]{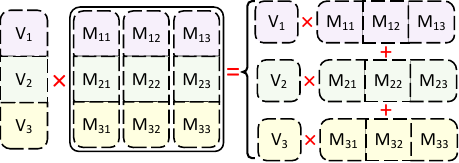}

\caption{Decomposing a vector-matrix multiplication into batches of scalar-vector operations, where each batch has a single scalar operand ($v_{1}, v_{2}, v_{3}$).}
\label{fig:vector_matrix_mult}
\vskip -0.15in
\end{figure}

Our approach consists of a lightweight mechanism that enables parallel column-independent accesses within each mat of a subarray. In SIMD operations where one operand is a scalar and thus it is the same for all the vector elements, Lama optimizes performance by requiring only a single ACT command for multiple LUT accesses. In this step, the scalar operand indexes the row page, while the vector elements enable simultaneous independent column accesses across different mats. This design is particularly effective for bulk multiplications involving a fixed operand, as commonly seen in vector-matrix and matrix-matrix multiplications, where the same input value is used in multiple operations. As illustrated in figure~\ref{fig:vector_matrix_mult}, a vector-matrix multiplication is broken down into several scalar-vector multiplication steps. A naive mechanism would require one LUT access for each individual multiplication, involving both an ACT command and a memory read command per operation. However, by grouping these operations into batches that share a scalar operand, all multiplications within the batch can be executed with a single ACT command followed by multiple memory read commands. This means that only the first LUT access requires an ACT command, which is then reused for the remaining LUT accesses within the batch. This approach reduces the need for total memory commands by $19.4x$ compared to a state-of-the-art work named pLUTo~\cite{pluto} for INT4 multiplications, thereby lowering overall latency and energy consumption.


Moreover, Lama draws inspiration from the mask logic circuit, found in commodity DRAM~\cite{micron_ddr4}, to support higher precision operations of up to 8-bit multiplications. Lama integrates easily with the existing commodity DRAM internal structure and organization, requiring minimal modifications with negligible (2.47\%) area overhead.

Finally, we introduce LamaAccel, an HBM-based PIM accelerator designed to enhance LLM inference by implementing our Lama PuM mechanism. LamaAccel utilizes a strategic data-mapping approach to allocate encoder-decoder blocks of LLMs across pseudo-channels within the HBM. This setup pipelines the processing of different layers within each encoder-decoder block, maximizing efficiency. LamaAccel integrates the DNA-TEQ~\cite{DNA-TEQ} scheme, an adaptive exponential quantization technique that simplifies complex multiply-and-add (MAC) operations into more efficient addition and counting operations (as detailed in Section~\ref{Exponential_quant}). This integration allows LamaAccel to support per-layer numerical precision of up to 8-bit, all while keeping accuracy loss below 1\%. This demonstrates the flexibility of the Lama technique in adapting to different arithmetic precisions, particularly in handling varying precision levels required by LLMs.

This paper makes the following key contributions:

\begin{itemize}

\item We present Lama, a novel lightweight memory access mechanism that enhances the efficiency of executing complex bulk arithmetic operations through a LUT-based PuM approach, exploiting the intrinsic mat-level parallelism and open-page policy of DRAM.


\item Lama supports multiplication operations with operand pairs of up to 8-bit precision and remains compatible with the standard DRAM timing parameters, introducing only a minimal area overhead of 2.47\%. We experimentally demonstrate that Lama significantly outperforms a PuM-based state-of-the-art approach~\cite{pluto} and a high-end CPU by $3.5\times$ and $3.8\times$ in performance and $8.3\times$ and $8\times$ in energy efficiency for bulk 8-bit multiplication tasks.

\item We propose LamaAccel, an HBM-based accelerator for attention-based models that implements our Lama PuM technique to perform addition and counting operations that replace traditional MAC operations. LamaAccel efficiently maps the encoder and decoder blocks of LLMs to different pseudo-channels within HBM, enabling parallel inference execution. We evaluate LamaAccel against a GPU, a TPU, and a state-of-the-art PuM-based technique~\cite{pluto} across various widely-used LLMs. The results show an average speedup of $4.1\times$$/7.2\times$ over TPU/GPU baselines and $1.7\times$ speedup compared to the state-of-the-art PuM technique. In terms of energy efficiency, LamaAccel achieves a $7.1\times$$/12\times$ energy reduction over TPU/GPU baselines, and $4\times$ energy savings compared to pLUTo.

\end{itemize}








\section{Background \& Related Work}
In this section, we review key terminology and concepts that are essential for understanding the content of this paper. First, we introduce the structure of HBM and its hierarchical micro-organization. Next, we provide an overview of attention-based models, focusing on the exponential quantization technique described in \cite{DNA-TEQ}. Finally, we examine prior PuM techniques~\cite{pluto,red-LUT}, discussing their main strengths and challenges.

\subsection{HBM Micro Organization}\label{HBM_organization}
\begin{figure}[t!]
\centering
\includegraphics[width=0.75\columnwidth]{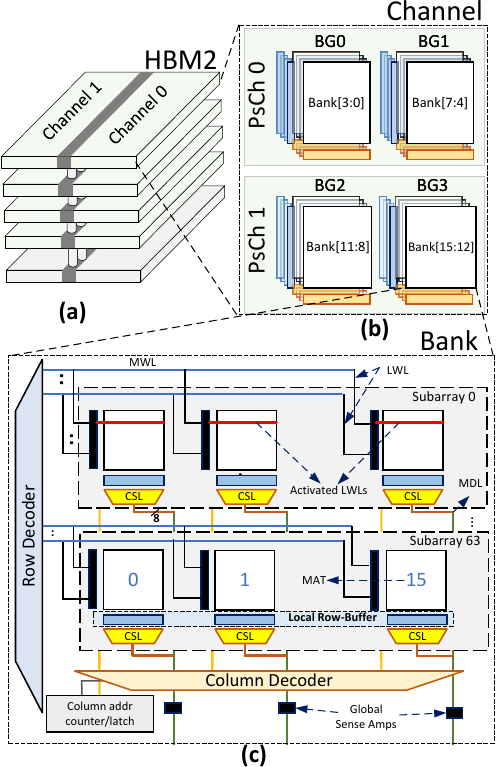}
\caption{HBM2 Organization: (a) High-level, (b) Channel-level, and (c) Bank-level hierarchy. The bank-level organization is adapted from \cite{fine-grained}.}
\label{fig:HBM_organization}
\vskip -0.15in
\end{figure}

The baseline memory architecture used in this work is based on micron's high-bandwidth memory (HBM2) as shown in Figure~\ref{fig:HBM_organization}a. An HBM stack consists of multiple DRAM slices stacked atop a base die, interconnected through numerous through-silicon vias (TSVs). This configuration delivers significantly higher bandwidth and lower access latency compared to conventional DRAM solutions.

Each DRAM die of HBM2 supports two independent channels and each 128-bit channel can be split into two separate 64-bit pseudo-channels~\cite{psch_mode}. In HBM2's pseudo-channel mode, the 64-bit I/O interface operates at a frequency of 1 GHz, offering a bandwidth of 16 GB/s over the DDR interface. The DRAM atom size, which is the smallest unit of data that can be accessed in a single request, is 32 bytes. This amount of data is transferred using a burst length of four over the 64-bit DDR I/O interface, ensuring efficient data throughput and reduced latency.

As shown in Figure~\ref{fig:HBM_organization}b, according to the JEDEC standard, a pseudo channel comprises two bank groups, with each group containing four banks. When operating in pseudo-channel mode, the two pseudo channels function semi-independently: they share the same channel’s row and column command bus as well as clock inputs, but decode and execute commands individually~\cite{jedec}. This mode enhances command bandwidth, reduces latency, and increases effective data bandwidth~\cite{demystifyingHBM}. Consequently, throughout this paper, we consider an HBM2 stack to be operating in pseudo-channel mode, with sixteen 64-bit wide channels per stack, rather than in legacy mode with eight 128-bit wide channels.

As shown in Figure~\ref{fig:HBM_organization}c, each DRAM bank is equipped with its own set of row and column decoders, along with a column-address counter/latch module~\cite{micron_ddr4}. The column-address counter/latch is responsible for selecting the specific column address within an activated row. Once a row is activated using an ACT command, the entire row's contents are loaded into the local row-buffer which serves as temporary storage. This allows subsequent read or write operations to access the data without needing to reactivate the row. According to \cite{fine-grained}, each read/write command must retrieve a 32-byte DRAM atom, accomplished through two \textit{internal column accesses}, with each access retrieving 16 bytes of data from 16 mats. For the first internal column access, the column counter/latch selects the column based on the address specified in the read/write command, while for the second internal column access, the column counter/latch increments the address to access the next set of data.



A modern DRAM bank is organized into a hierarchical structure consisting of \textit{subarrays} and \textit{mats}. Each subarray within a DRAM bank contains a subset of the \textit{rows} and is equipped with a set of sense amplifiers that form the local row buffer for that subarray. In an HBM2 Micron memory configuration, a bank of 32KB is organized into 64 subarrays, each comprising 512 rows of DRAM cells. To manage the long bitlines and wordlines effectively, each subarray is further divided into multiple quadratic units called \textit{mats}, each sized at $512\times512$. When a row is activated in a subarray via an ACT command, each mat within the subarray activates a segment of the row into its local set of sense amplifiers. The subarray performs this operation by first driving a Master Wordline (MWL) in a high-level metal across the entire subarray, which then activates the Local Wordlines (LWLs) in each individual mat. Subsequently, when a DRAM atom is read, every mat of the subarray outputs a few bits of the DRAM atom~\cite{fineDRAM}.

In HBM2, a row with a width of 1KB, referred to as the \textit{page size}, is distributed across 16 mats, each 512-bit wide. During a read operation, each mat provides 8-bit per internal cycle. Over two internal cycles, all mats collectively deliver the 32-byte DRAM atom~\cite{fineDRAM}. The read command activates the Column-selection Logic (CSL) in each mat, which functions as a multiplexer, directing data from the target sense amplifiers to the master data lines (MDLs) through the Local Data Lines (LDLs)~\cite{drambook, understandingDRAM}. The bank I/O logic, or global sense amplifiers, connects to all mats via the MDLs, facilitating the aggregation and transmission of data from the individual mats to the external data bus.

\subsection{Attention-based models}
\begin{figure}[t!]
\centering
\includegraphics[width=0.9\columnwidth]{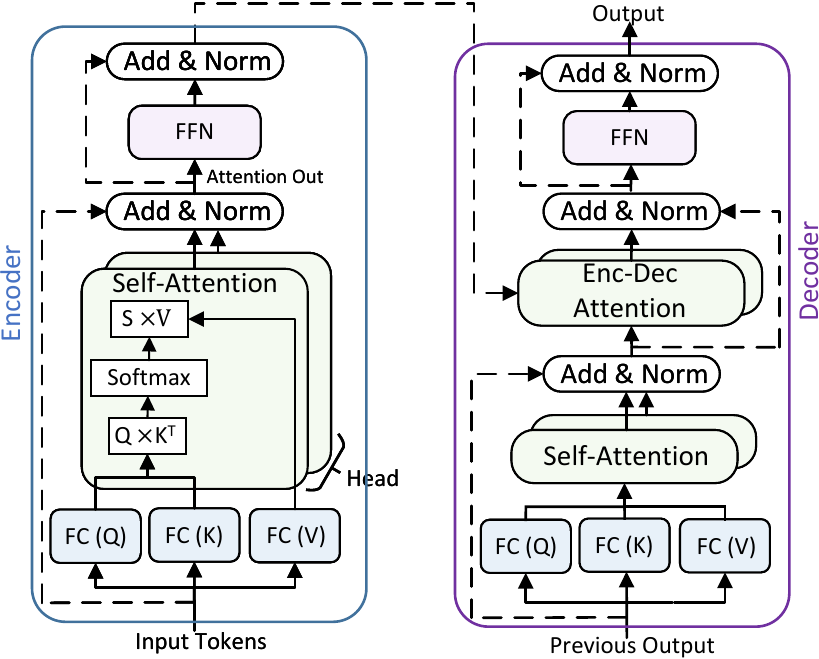}
\caption{Encoder and Decoder blocks in Attention-based models.}
\label{fig:Transformers}
\vskip -0.15in
\end{figure}

Attention-based models, also known as Transformers, are composed of a stack of encoders, decoders, or a combination of both. As shown in Figure~\ref{fig:Transformers}, each encoder includes a self-attention layer followed by a feed-forward network (FFN).

For an input sequence with \textit{L} tokens, the encoder attention block generates a query vector $Q$, a key vector $K$, and a value vector $V$ per token. This is done by passing each input token through three fully-connected (FC) layers, where each layer multiplies the input by learned weight matrices $W^Q, W^K, W^V$, respectively. Afterward, the query matrix and the key matrix are multiplied together to compute the attention scores, which indicate how well each query vector correlates with the corresponding key vectors. Softmax is then applied to the resulting matrix to normalize the scores, converting them into probabilities that sum to one. This step is often followed by a dropout to prevent overfitting, yielding to the self-attention matrix $S$. The attention matrix $S$ is then multiplied by the value matrix $V$ to produce the final output of the attention mechanism. The self-attention layer is usually split into multiple heads, processed in parallel, where each head pays attention to a different region of the input sequence. The results from each attention head are then concatenated into a single matrix and merged through a hidden fully-connected layer. Finally, the add and norm layer applies a residual connection followed by layer normalization. The residual connection adds the input of the attention layer to its output, while the normalization layer ensures that the output values remain within a specific range. At the end of the encoder block, the attention output is passed through a FFN, consisting of two fully-connected layers that generate the block’s final output.


The produced output can then serve as input to the next block (either an encoder or decoder) or be directed to a task-specific output layer, such as for classification. In text generation tasks decoder blocks follow a similar structure and flow as encoder blocks but with some key differences. Unlike encoder blocks, which handle sequences of tokens, the input and output of decoder blocks typically consist of only one or a few tokens at a time. Additionally, decoder blocks include a masked self-attention layer that ensures predictions are based only on previously generated tokens, preserving the autoregressive property essential for tasks like language generation. In text generation tasks, there is one extra layer called encoder-decoder attention which allows the decoder to focus on specific parts of the encoded input sequence while generating the output. It uses the decoder's query and the encoder's key-value pairs to compute attention scores. Overall, attention-based models involve numerous memory-intensive and matrix multiplication operations~\cite{transpim, A3, mnnfast, interstellar}, making them well-suited for in-memory and near-memory acceleration.

\vspace{-2.5mm}

\subsection{Exponential Quantization}\label{Exponential_quant}
Researchers have proposed various optimizations for deep neural networks (DNNs), including pruning, clustering, and quantization~\cite{permdnn}. These techniques aim to reduce the memory footprint and computational complexity of DNNs, facilitating their deployment in real-time applications. Uniform quantization~\cite{ReDY} has been widely utilized to reduce the numerical precision (usually to 8-16 bits) and computational cost, typically with minimal impact on accuracy. To further decrease the model size and computational overhead, a recent study \cite{mokey} introduced techniques that assign varying bitwidths to weights and/or activations based on their sensitivity to quantization error. This method analyzes the distribution of each tensor, using it as a criterion to determine the optimal numerical precision for each DNN layer.

Similarly, DNA-TEQ~\cite{DNA-TEQ} employs a mixed-precision scheme for each layer, but with a non-uniform quantization representation that minimizes accuracy loss by reducing quantization error. DNA-TEQ leverages exponential quantization, which exploits the property $b^{a} \cdot b^{w} = b^{a + w}$. This approach not only reduces memory consumption but also simplifies the hardware implementation by replacing expensive multiplication operations with cost-effective addition operations. DNA-TEQ quantizes values into the form $S_{W}(\alpha_{W} b^{int_{W}} + \beta_{W})$ for weights $W$ and $S_{A}(\alpha_{A} b^{int_{A}} + \beta_{A})$ for activations $A$, where $S_{W}$, $S_{A}$ represent the sign, $int_{W}$, $int_{A}$ are signed $n$-bit integer exponents, $\alpha_{W}$, $\alpha_{A}$ are scale factors and $\beta_{W}$, $\beta_{A}$ are offsets for the respective weight and activation tensors.

As shown in Equation~\ref{eqn:conv_extended}, each output activation is computed as a dot-product of activations and weights, which can be expanded into a sum of four simpler terms:

\begin{equation}
\begin{scriptsize}
\begin{aligned}
\label{eqn:conv_extended}
\sum_{i=1}^{m} A_{i} \cdot W_{i} = \sum_{i=1}^{m}S_{A_{i}} (\alpha_{A} b^{int_{A_{i}}} + \beta_{A}) \cdot S_{W_{i}} (\alpha_{W} b^{int_{W_{i}}} + \beta_{W}) =\\
\underbrace{\alpha_{A}\alpha_{W}\sum_{i=1}^{m} (S_{A_{i}} S_{W_{i}})b^{int_{A_{i}} + int_{W_{i}}}}_\text{1} + \underbrace{\alpha_{W}\beta_{A}\sum_{i=1}^{m}(S_{A_{i}} S_{W_{i}})b^{int_{W_{i}}}}_\text{2} \\
+ \underbrace{\alpha_{A}\beta_{W}\sum_{i=1}^{m}( S_{A_{i}} S_{W_{i}}) b^{int_{A_{i}}}}_\text{3} + \underbrace{\beta_{A}\beta_{W}\sum_{i=1}^{m}S_{A_{i}} S_{W_{i}}}_\text{4}
\end{aligned}
\end{scriptsize}
\end{equation}

The first term requires the addition of exponents, denoted as $b^{int_{A_{i}} + int_{W_{i}}}$, and can be computed by counting the number of times that each addition of exponents occurs. The second term involves the summation of base $b$ to the power of weight exponents, adjusted by the sign of both the weight and activation. Similarly, the third term focuses on the summation of $b$ to the power of activation exponents, adjusted by their corresponding signs. Both the second and third terms can be computed by counting the occurrences of the exponents. Finally, the fourth term accumulates the sign products, which can be derived from any of the previous terms by summing the total number of occurrences.

After counting the occurrences associated with each exponent, these counts are multiplied by their corresponding values $b^{int}$. The resulting products are then accumulated into a single value, which is subsequently multiplied by the constant coefficients, and all terms are summed together to produce the final output activation.

\subsection{LUT-based PuM Techniques}\label{pluto_background}
\begin{figure}[t!]
\centering
\includegraphics[width=1.0\columnwidth]{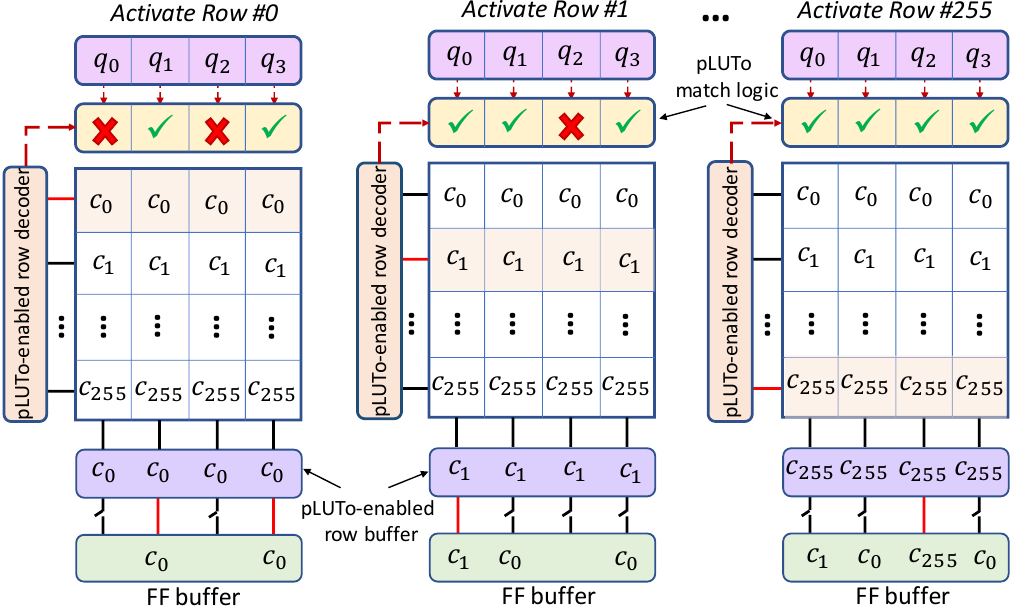}
\caption{Example of pLUTo for 4-bit multiplications.}
\label{fig:pLUTo}
\vskip -0.15in
\end{figure}

The authors of pLUTo \cite{pluto} proposed a novel DRAM-based PuM approach that employs LUTs to perform complex operations efficiently. As illustrated in Figure~\ref{fig:pLUTo}, pLUTo can be used to perform a 4-bit bulk multiplication task with LUTs.

In this example, the LUT is designed to store pre-computed multiplication results for all possible 4-bit operand combinations. When performing the multiplication of two 4-bit operands, $a$ and $b$, where $a = \overline{a_{3}a_{2}a_{1}a_{0}}$ and $b = \overline{b_{3}b_{2}b_{1}b_{0}}$, the result $c$, is an 8-bit value. To compute $c = a\times b$ using pLUTo, the LUT query input vector is formed by concatenating the binary representations of $a$ and $b$. The input bit vector $q_{x}=[a,b]$ is then used to query the LUT, which returns the results of four multiplication queries $q_{0}$ to $q_{3}$ in a single operation. Although the LUT can accommodate up to one whole row of queries, for simplicity, only four queries are depicted in the figure. This allows for efficient multiplication directly within the memory.

To perform four LUT queries simultaneously, pLUTo stores and replicates the LUT four times within a single DRAM subarray. As shown in Figure~\ref{fig:pLUTo}, each row $i$ in the subarray contains repeated copies of the element corresponding to the $i$-th index of the LUT. The memory controller first loads the query input vector into the source row buffer. Then, it initiates a \textit{pLUTo Row Sweep} operation, which sequentially activates all rows in the pLUTo-enabled subarray that hold the LUT elements. The activation follows an orderly pattern, such as activating rows $\#0$, $\#1$, and so on, up to row $\#255$.

During each row activation, the \textit{pLUTo Match Logic} identifies matches between the current row index in the subarray and the elements in the LUT query input vector. This matching process is essential for selecting the correct elements from the LUT that correspond to the query. Once a match is detected, the relevant elements in the pLUTo-enabled row buffer are copied into the \textit{Flip-Flop buffer}, which stores the final output for the whole LUT query operation.

pLUTo faces several challenges when performing calculations using LUTs within DRAM subarrays. First, the \textit{pLUTo Row Sweep} operation, which activates rows sequentially, results in long latency and high energy consumption due to repeated ACT commands. Second, pLUTo's ability to perform parallel LUT queries within a bank is limited by DRAM’s tFAW timing constraints. In addition, the technique introduces a 16\% area overhead due to the modifications to the row decoder and the inclusion of the match logic and extra buffer. Finally, scaling pLUTo for operations with more than 4-bit precision is costly, as it supports a maximum of 8-bit LUT query inputs. To handle higher precisions, operations must be divided into smaller segments and processed across multiple subarrays, adding complexity. Lama tackles all these challenges and provides important energy savings over pLUTo as we will show later in this paper.

\section{Lama Overview}\label{LamaOverview}
Lama is a lightweight LUT-based mechanism designed to efficiently execute complex arithmetic operations in bulk. Our primary goal is to address the critical inefficiencies of existing LUT-based PuM approaches, particularly those caused by the need for successive ACT commands during bulk operations, as well as the challenges associated with supporting operands larger than 4 bits. Additionally, Lama tackles the performance limitations imposed by DRAM's structural constraints. To overcome these challenges, Lama introduces a novel execution scheme that eliminates the reliance on successive ACT commands for performing bulk arithmetic operations. As shown previously in Figure~\ref{fig:vector_matrix_mult}, Lama clusters operations into batches, which are defined for simpler use as \textit{operand-coalesced batches}, where each batch consists of multiple operations that share a scalar operand. In other words, an operand-coalesced batch is a function $f$ whose inputs are a scalar $a$ and a vector $b$, and the result is another vector that consists of applying the function $f$ to each pair of elements $a$ and $b_{i}$.

To perform LUT-based computing and derive the result of a function $f$ applied to operands $a$ and $b$ (i.e., $f(a,b)$), Lama defines two key operations: \textbf{\textit{LUT activation}} which involves a memory ACT operation that activates a row, with the row index determined by the value of $a$, and \textbf{\textit{LUT retrieval}}, where the actual results for concurrent operations $f(a,b_{i})$ are fetched through one internal column access (ICA). The starting point for the column access is determined by the value of $b_{i}$, which addresses the specific column positions in the activated row. These two operations are executed consecutively to efficiently compute the desired function result.

Lama supports LUT retrieval for any given function $f$, enabling arithmetic operations on operands with bit-widths of up to 8-bit. Depending on the specific operation, the result $f(a,b_{i})$ can have a bit-width of up to 16-bit (e.g., multiplication).

\begin{figure}[t!]
\centering
\includegraphics[height=2.7cm, width=8.9cm]{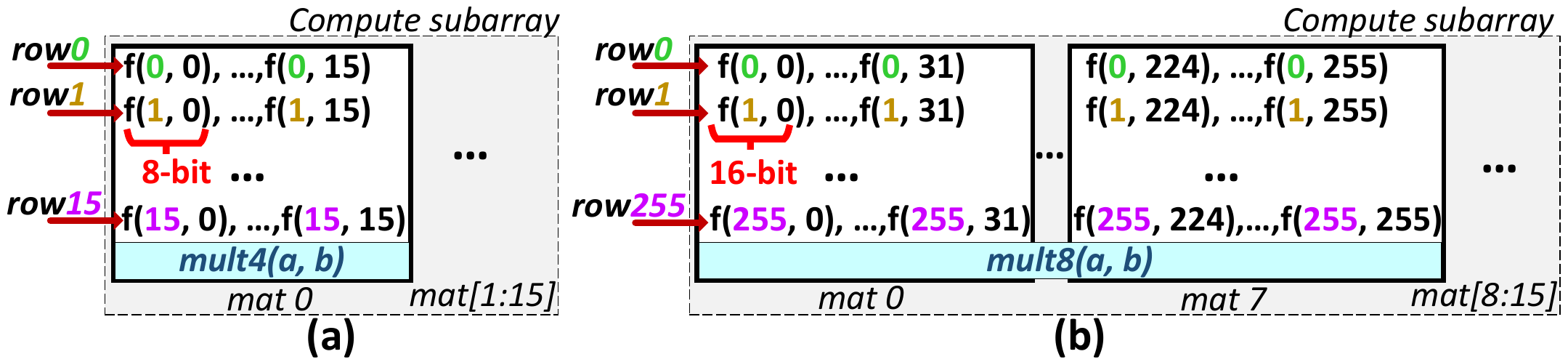}
\caption{LUT data layout for (a) 4-bit and (b) 8-bit multiplications.}
\label{fig:lut_data_layout}
\vskip -0.15in
\end{figure}

Figure~\ref{fig:lut_data_layout} illustrates the data layout for a set of LUTs used to perform bulk operations on operand-coalesced batches. Each LUT is dedicated to one independent function $f(a,b_{i})$. The size of the LUT varies based on the operands and result precision of function $f$. For example, a 4-bit multiplication LUT occupies one mat, while an 8-bit multiplication LUT requires eight mats to fit its data. To enable parallel processing, the LUT is replicated across the entire subarray. The \textit{degree of parallelism}, denoted as $p$, reflects the number of simultaneous operations that can be performed through a LUT retrieval, and is influenced by the LUT data size. In scenarios where the LUT data size is small, such as 4-bit multiplications, $p$ can reach up to 16, i.e., the number of mats available in HBM2. On the other hand, for larger LUT data sizes like 8-bit multiplications, $p$ is reduced to 2 due to the increased data requirements.

If the number of parallel operations ($p$) that can be executed simultaneously is less than the size of the operand-coalesced batch, completing all operations within the batch requires issuing additional memory commands. However, Lama avoids the need for extra LUT activations in this scenario. The reason is that the LUT activation, which is based on the scalar operand, remains valid for the entire operand-coalesced batch. Consequently, Lama reuses the already activated row in the local row buffer by taking advantage of the open-page policy. Only the LUT retrievals for the subsequent sets of $p$ operations are performed, ensuring that only a single ACT command is required for the entire batch of operand-coalesced operations.

\subsection{Enabling Independent Column Selection in DRAM Mats}
In conventional commodity DRAMs, memory read operations involve a column-address counter/latch that selects the same column position across all mats within a subarray using Column Select Lines (CSLs). Since each mat provides 8-bit data per column access, the column counter can select from 64 different data positions in HBM2. However, to efficiently execute operand-coalesced batches in bulk, independent column access across mats is essential. This capability allows each mat to access different data positions within the same row.

To facilitate this, Lama replicates the column counter/latch within each bank to match the number of mats in a subarray. In HBM2, where each subarray consists of 16 mats, Lama introduces 16 column counters/latches per bank, enabling independent column selection across all mats. This setup is suitable because only one subarray can connect its local data lines to the global sense amplifiers at a time for read or write operations, and thus there is no need to replicate per subarray.

Moreover, to accommodate both conventional memory operations and LUT-based modes, Lama's column counters are designed to receive the column selection signal from the address register or from a temporary buffer via a multiplexer. The added area overhead of these replicated column counters is minimal, and their impact on overall memory capacity and performance is negligible, as will be detailed in Section~\ref{overhead}.

\subsection{Mask Logic}
When performing a LUT retrieval operation, each mat transfers consecutive column positions, but not all of the fetched data from mats may correspond to the desired result $f(a,b_{i})$. This issue arises when the LUT data for the function $f$ spans across multiple mats, reducing the degree of parallelism ($p$) to less than 16. In such cases, the mask logic selects valid results while filtering out irrelevant data from the remaining mats.

\begin{figure}[t!]
\centering
\includegraphics[height=1.8cm, width=4cm]{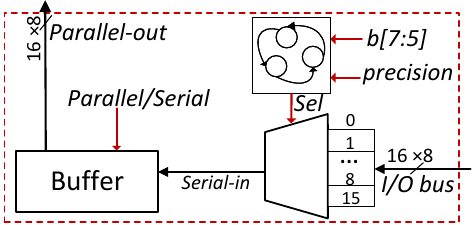}
\caption{Mask logic.}
\label{fig:mask_logic}
\end{figure}

The mask logic's functionality is illustrated in Figure~\ref{fig:mask_logic}. The design features a multiplexer with sixteen inputs, each connected to a different mat in the subarray. To filter out irrelevant data and retain the target values, the mask logic operates in a serial manner, where in each iteration, the select port of the multiplexer is determined by a small finite-state machine that uses the most significant bits (MSBs) of the vector element $b_{i}$ and the precision of the operation to choose the correct value.


Because the number of valid results $f(a,b)$ retrieved in a single LUT access depends on the precision and degree of parallelism, the size of the result can vary. To handle this variation, a buffer is placed after the multiplexer, concatenating results until they reach 16 bytes before outputting. This ensures a consistent bitwidth across different precision levels and parallelism degrees.


\section{Case study 1: Lama for Bulk Multiplications}
In this section, we detail how the Lama technique handles bulk multiplications for one operand-coalesced batch. To simplify the explanation, we first describe the process for 4-bit bulk multiplications and then discuss the differences when scaling up to 8-bit bulk multiplications.

\begin{figure}[t!]
\centering
\includegraphics[width=1.0\columnwidth]{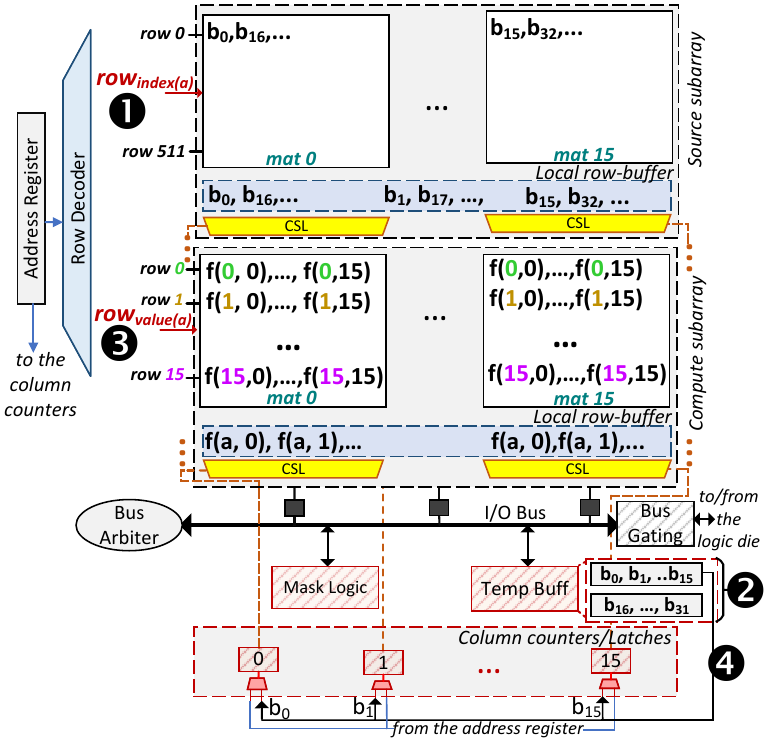}
\caption{Detailed steps to perform bulk 4-bit multiplications using Lama.}
\label{fig:case_study1}
\vskip -0.15in
\end{figure}

As outlined in Section~\ref{HBM_organization}, each memory bank in DRAM is organized into subarrays, with each subarray consisting of 16 mats. Lama leverages both bank-level parallelism and mat-level parallelism within each bank, using all 16 mats in a subarray to perform simultaneous arithmetic operations, thereby enabling efficient bulk execution. Figure~\ref{fig:case_study1} illustrates the DRAM bank structure with the modifications of Lama to carry out bulk multiplications. To allow multiple subarrays to keep their rows open simultaneously, the global row buffer must be isolated from the local row buffers within each subarray. To achieve this, Lama employs a technique proposed in \cite{salp}, which involves adding a tri-state buffer per mat after the column-selection logic (CSL). This prevents short circuits in the master data line (MDL) for activated mats aligned in the same vertical position. This method ensures that multiple rows in different subarrays can remain open, while only one \textit{designated} subarray is used to handle column commands at any given time.


Figure~\ref{fig:case_study1} provides an overview of the DRAM bank structure and the execution flow required to implement Lama. Lama involves several key components: the \textbf{source subarray}, which stores the vector operands $b$ for all operand-coalesced batches, with the elements in each vector organized across one or more rows; the \textbf{compute subarray}, which holds the LUT data following the layout shown in Figure~\ref{fig:lut_data_layout} (for 4-bit operations each LUT for $f(a,b_{i})$ occupies a single mat and it is replicated in 16 mats); the \textbf{temporary buffer}, a small buffer used to temporarily store elements of the vector operand $b$ fetched from the source subarray; the \textbf{mask logic}, which filters and processes fetched data as needed; the \textbf{column counters}, previously described in Section \ref{LamaOverview}; and finally, the \textbf{bus arbiter}, responsible for managing data transfers through the global I/O bus and coordinating interactions between the mask logic, temporary buffer, and global row buffer.

In addition, for simplicity, we refer to the process of reading elements of the vector operand $b$ from the source subarray and storing them in the temporary buffer as an \textit{internal memory read} operation. Each internal read consists of two consecutive \textbf{internal column accesses (ICAs)} that fetch the entire 32-byte DRAM atom, corresponding to 8 bits per mat in each ICA. It is important to note that the key distinction between an internal read and a standard memory read is that the internal read does not transfer data to the I/O buffer nor sends it to the host; instead, it stores the data locally in the temporary buffer.


As previously mentioned, Lama decomposes the vector-matrix multiplication into multiple operand-coalesced batches. Each batch corresponds to a scalar-vector operation with a scalar operand $a$ and a vector operand $b$, where the elements of $b$ are stored across one or more rows in the memory. To manage row addressing in the \textit{source subarray}, Lama relies on the \textit{positional index} of the scalar operand $a$ within its original vector before decomposing the vector-matrix multiplication. This positional index is used to issue the ACT command to the corresponding row in the source subarray, which holds the coalesced batch related to that scalar operand. If the vector operand $b$ of the coalesced batch exceeds the storage capacity of a single row, Lama scales the positional index to span multiple rows. In contrast, for the \textit{compute subarray}, the \textit{value} of $a$ is used to ACTIVATE the appropriate LUT row where the function results are stored. Further details on this activation process are provided later in the section.


Lama execution flow employs a systematic approach to perform bulk multiplications for one of the operand-coalesced batches residing in the source subarray. Here we explain the process for 4-bit bulk multiplications as an example. First, in the source subarray, a row is activated based on the corresponding positional index of the scalar operand $a$ through an ACT command (\circled{1}), and all elements of vector operand $b$ in the selected row are stored in the local-row buffer of the source subarray, where they will remain for multiple iterations. Then, an internal read command is issued to fetch 32 different elements of $b$ and store them in the temporary buffer (\circled{2}). It is important to note that Lama statically performs zero-padding on elements of $b$ to match the fixed bitwidth of the hardware no matter the precision, that is, 4-bit elements are padded to 8-bit, which corresponds to the reading granularity of each mat. The indexed row in the source subarray is left open after the internal read operation, and remains open until all elements of $b$ in that row have been processed to finish the computations of the operand-coalesced batch.


Next, the LUT-based computation begins in the compute subarray by performing a LUT activation based on the \textit{value} of the scalar $a$ as an index (\circled{3}). After the desired row is activated and resides in the local row buffer, the column address counter/latch units initiate the LUT retrieval operation using the first 16 elements of $b$ stored in the temporary buffer ($b_{0}$ to $b_{15}$) as shown in\circled{4}. In the first LUT retrieval, each mat outputs an 8-bit result for the operation $f(a,b_{i})$. These results are directly transferred to the host via the global I/O bus. For the next set of 16 elements of $b$, the column counters perform an additional LUT retrieval, and the results are again transferred to the host.

This iterative process continues until all operations for the entire operand-coalesced batch are completed. Once all computations are finalized, a PRECHARGE command is issued to close the open rows in both the source and compute subarrays. In the event that the elements of $b$ span over several rows, additional ACT and PRECHARGE commands are issued per row, requiring more iterations to complete the batch. Note that with a parallelism degree of $p = 16$, all LUT-retrieved values are valid, leaving the mask logic idle throughout the entire process for 4-bit bulk multiplications.


\subsection{8-bit Multiplications}
\begin{figure*}[t!]
\centering
\includegraphics[width=1.0\textwidth]{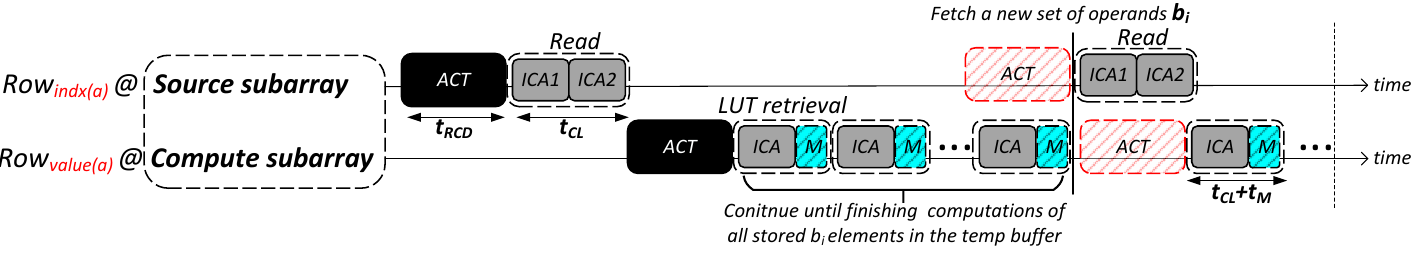}
\caption{Timeline of performing multiplications using Lama. Each read operation includes two internal column accesses (ICA). The "M" block represents the latency of the mask logic. LUT retrieval includes one ICA followed by the mask logic. An ACT command is issued only once to access the row indexed by the scalar operand $a$ in both the source and compute subarrays. For subsequent iterations with new elements of $b$, no additional ACT commands are needed (indicated by the red block) due to the open-page policy.}
\label{fig:timeline_multiplication}
\vskip -0.15in
\end{figure*}

Figure~\ref{fig:timeline_multiplication} depicts the timeline of performing 8-bit bulk multiplications within a memory bank using Lama. The process for accessing elements of $b$ from the source subarray remains consistent with the 4-bit operation, where Lama fetches 32 different $b$ elements per internal memory read. However, the LUT-based computation differs in two key aspects when scaling up to 8-bit operations.

First, the LUT data for an 8-bit multiplication spans across eight mats (Figure~\ref{fig:lut_data_layout}b), reducing the parallelism degree to $p=2$. During one LUT retrieval, only two elements of $b$ are selected from the temporary buffer and broadcasted to all 16 column counters, with each set of eight consecutive column counters processing the same element $b$. While this procedure limits the parallelism within each bank, Lama compensates it by leveraging bank-level parallelism, maintaining high throughput even when handling higher operand precision.


Second, after performing the LUT retrieval for these two elements of $b$, not all fetched values are valid. Consequently, the mask logic filters out invalid results within each set of eight mats, by using the three MSBs of $b$ elements to select the mat containing the valid result of each set. The valid results are stored in the mask logic buffer until gathering 16 bytes, and then they are outputted. The mask logic operates in two cycles (since $p=2$) to serially select the valid $f(a,b_{i})$ for each element of $b$, and its latency is added to the LUT retrieval operation as illustrated in Figure~\ref{fig:timeline_multiplication} by the "M" block. As demonstrated in the evaluation section~\ref{eval_mult8b}, this added latency hardly impacts performance compared to state-of-the-art PuM techniques.

To complete the fetching of a 16-bit multiplication result for each of the two elements of $b$, two LUT retrieval operations are done. This is because each internal column access (ICA) retrieves 8 bits from each mat, and only two of the 16 mats provide valid data. During the first LUT retrieval, data is accessed according to the two elements of $b$. For the second LUT retrieval, the column counters increment the previously determined addresses by one, to access the subsequent 8 bits in each mat.

This procedure continues until all computations related to the current operand-coalesced batch are completed. However, the ACT command in both the source and compute subarray is only issued once during the initial iteration. For all subsequent iterations, Lama efficiently reuses the already-opened row in the local row buffer of both subarrays, eliminating the need for additional ACT commands. Additional memory commands are only necessary when the $b$ elements are stored in multiple rows of the source subarray.


Lama is not limited to multiplication operations; it can execute any arithmetic function $f(a,b)$, including addition, division, and other complex functions. Multiplication is used as an example in this context due to its relevance, and the computational complexity it introduces when handled within memory systems.

\subsection{Column Addressing for Varying Operand Precision}
For all supported operand precisions except 4-bit, the LUT multiplication results are extended to a 16-bit format to be word aligned in memory, requiring two Internal Column Accesses (ICAs) to fetch the complete result. During each ICA, the column counter selects from 64 possible positions (each mat has 64 8-bit columns), using the elements of $b$ to determine the correct address. The 5 least significant bits (LSBs) of $b$ with a zero bit appended to the right,\{$b_{i}[4:0], 0$\}, is used as the 6-bit column address for the first ICA. For the second ICA, the appended bit is changed to one to fetch the next 8 bits from each mat to complete the 16-bit result. For 5-bit precision, all 16 mats provide a valid result (the LUT occupies a single mat and is replicated 16 times in each bank), each mat providing the result for a different $b_{i}$. For precisions higher than 5 bits, the LUT occupies multiple mats, so the remaining most significant bits (MSBs) of $b_{i}$ are used by the mask logic to select the valid results. For instance, for 6-bit precision, the LUT occupies 2 mats, and the most significant bit of $b_{i}$ is used to select which one of each pair of mats provides the desired result. In this case, each bank can process 8 operations in parallel, each corresponding to a different element of $b$.

Table~\ref{t:mult_different_precisions} shows the degree of parallelism per bank ($p$) and the number of bits from $b$ required for column addressing across varying operand precisions. For 4-bit and 5-bit multiplications, the parallelism level $p$ reaches its maximum value, utilizing all 16 mats within a subarray. The mask logic is bypassed in these cases, as all fetched results are valid. In 6-bit multiplication, $p$ is reduced to 8, as the LUT data spans two mats. The 5 LSBs of $b_i$ are used for column selection, while the MSB is used by the mask logic to select one data element out of two consecutive mats. In 7-bit multiplication, the LUT data spans four mats, decreasing $p$ to 4. In this case, the two MSBs of $b_i$ are used by the mask logic to select a valid mat among four consecutive mats. For 8-bit multiplication, the three MSBs select one element among eight consecutive mats.

\begin{table}[t!]
\caption{Parallelism Degree ($p$), Column Addressing, and Masking Requirements for Different Operand Bitwidths in Lama MUL.}
\label{t:mult_different_precisions}
\vskip -0.20in
\begin{center}
\resizebox{1.0\columnwidth}{!}{%
    \begin{tabular}{|c|c|c|c|c|}
    \hline
    \textbf{bitwidth}                  & \textbf{p}                 & \textbf{\#LSBs for Column Addressing}        & \textbf{\#MSBs for Masking}          & \textbf{\#ICAs}  \\
    \hline
               4-bit                   &   16                       & 4                                                            &  $\times$                                            & 1                \\
    \hline
               5-bit                   &   16                       & 5                                                            &  $\times$                                            & 2                \\
    \hline
               6-bit                   &   8                        & 5                                                            &  1                                                   & 2                \\
    \hline
               7-bit                   &   4                        & 5                                                            &  2                                                   & 2                \\
    \hline
               8-bit                   &   2                        & 5                                                            &  3                                                   & 2                \\
    \hline
    \end{tabular}%
}
\end{center}
\vskip -0.20in
\end{table}

\subsection{Column Addressing for Varying Precision}
For operand precisions greater than 4 bits, LUT multiplication results are extended to a 16-bit format to be word-aligned in memory, requiring two ICAs to fetch the complete result. During each ICA, the column counter selects from 64 possible positions (each mat has 64 8-bit columns), using the elements of $b$ to determine the correct address. The 5 least significant bits (LSBs) of $b$ with a zero bit appended to the right, \{$b_{i}[4:0], 0$\}, serve as the 6-bit column address for the first ICA. For the second ICA, the appended bit is changed to one to fetch the next 8 bits from each mat, completing the 16-bit result. For 5-bit precision, all 16 mats provide a valid result (the LUT data occupies a single mat and is replicated 16 times in each bank), each mat providing the result for a different $b_{i}$. For precisions higher than 5 bits, the LUT spans multiple mats, so the remaining MSBs of $b_{i}$ are used by the mask logic to select the valid results. For instance, in 6-bit precision, the LUT occupies 2 mats, and the MSB of $b_{i}$ selects which one of each pair of mats provides the desired result. In this case, each bank can process 8 operations in parallel, each corresponding to a different element of $b$.

Table~\ref{t:mult_different_precisions} summarizes the degree of parallelism per bank ($p$) and the number of bits from $b$ required for column addressing across varying operand precisions. For 4-bit and 5-bit multiplications, $p$ reaches its maximum value, utilizing all 16 mats within a subarray, with the mask logic bypassed since all fetched results are valid. In 6-bit multiplication, $p$ is reduced to 8, as the LUT data spans two mats, using the 5 LSBs of $b_i$ for column selection and the MSB by the mask logic to select one data element out of two consecutive mats. In 7-bit multiplication, the LUT data spans four mats, decreasing $p$ to 4, with the two MSBs of $b_i$ are used by the mask logic to select a valid mat among four consecutive mats. For 8-bit multiplication, the three MSBs select one element among eight consecutive mats.

\begin{table}[t!]
\caption{Parameters for different operand bitwidths in Lama MUL.}
\label{t:mult_different_precisions}
\begin{center}
\resizebox{1.0\columnwidth}{!}{%
    \begin{tabular}{|c|c|c|c|c|}
    \hline
    \textbf{bitwidth}                  & \textbf{p}                 & \textbf{\# $b_{i}$ LSBs for Column Addressing}        & \textbf{\# $b_{i}$ MSBs for Masking}          & \textbf{\#ICAs}  \\
    \hline
               4-bit                   &   16                       & 4                                                            &  $\times$                                            & 1                \\
    \hline
               5-bit                   &   16                       & 5                                                            &  $\times$                                            & 2                \\
    \hline
               6-bit                   &   8                        & 5                                                            &  1                                                   & 2                \\
    \hline
               7-bit                   &   4                        & 5                                                            &  2                                                   & 2                \\
    \hline
               8-bit                   &   2                        & 5                                                            &  3                                                   & 2                \\
    \hline
    \end{tabular}%
}
\end{center}
\vskip -0.1in
\end{table}

\subsection{Methodology and Configuration}
This section presents the methodology for evaluating Lama's performance and energy consumption in the case study of bulk multiplications. Table~\ref{t:configuration} outlines the hardware characteristics of Lama's evaluation platform. Our implementation assumes a memory setup based on the standard HBM2 specification~\cite{jedec}. The timing and energy parameters are derived from \cite{fine-grained}.

As shown in Table~\ref{t:configuration}, the evaluation of Lama also considers bank-level parallelism by distributing different coalesced batches across various banks. As shown in Figure~\ref{fig:timeline_multiplication}, performing all computations for a coalesced batch requires only two ACT commands. Consequently, when all banks within a channel are utilized for computation, a total of 32 ACT commands are issued. To accommodate these 32 ACT commands, 8 {\fontfamily{cmss}\selectfont tFAW} windows are needed. To avoid any stalls while all banks are operational, the computation time for each coalesced batch must exceed 8 times the {\fontfamily{cmss}\selectfont tFAW} window duration. For 4-bit precision, which has the shortest computation time, the coalesced batch size must be greater than 128 elements of $b$ to avoid being limited by the {\fontfamily{cmss}\selectfont tFAW} constraint. For other precision, this restriction is less stringent since the computation time per batch is longer.

Furthermore, within each bank, only one subarray is required and enabled to perform arithmetic operations for a given function, including those with 8-bit precision operands. To accommodate various functions, additional subarrays can be allocated for LUT computation. The host-to-HBM bandwidth is assumed to be 256 GB/s, consistent with the specifications in \cite{fine-grained}.

\begin{table}[t!]
\caption{Architectural Parameters for Lama.}
\label{t:configuration}
\begin{center}
\resizebox{1.0\columnwidth}{!}{%
    \begin{tabular}{|>{\centering\arraybackslash}m{3.3cm}|m{7.4cm}|}
    \hline
    \makecell{\textbf{HBM Organization}}  &    channels/die (4-die stack) = 2 (8), pch/channel = 2,\newline  banks/channel = 16  (banks/pch = 8),  banks/group = 4,  \newline subarrays/bank = 64, bank rows = 32k, row buffer size/channel (row buffer size/pch)= 2KB (1KB), mat size = $512\times512$, DQ size = 128-bit/channel            \\
    \hline
    \textbf{HBM Timing (ns)}              &  t\textsubscript{RC} = 45, t\textsubscript{RCD} = 16, t\textsubscript{RAS} = 29, t\textsubscript{CL} = 16, t\textsubscript{RRD} = 2, t\textsubscript{WR} = 16, \newline t\textsubscript{CCD\textsubscript{S}} = 2, t\textsubscript{CCD\textsubscript{L}} = 4, t\textsubscript{FAW} = 12,  \# of activates in t\textsubscript{FAW} = 8  \\
    \hline
    \textbf{HBM Energy (pJ)}              &    e\textsubscript{ACT} = 909, e\textsubscript{Pre-GSA} = 1.51, e\textsubscript{Post-GSA} = 1.17, e\textsubscript{I/O} = 0.80  \\
    \hline
    \textbf{Bank-level Configuration}              &        Clock = 500MHz, Column Counter/Latch = $16(8b)$,  \newline  Mask Logic = $1$, Temporary Buffer = $1(64B)$               \\
    \hline
    \end{tabular}%
}
\end{center}
\end{table}

The objective of our evaluation is to demonstrate that Lama significantly reduces the number of memory commands, particularly activation (ACT) commands, compared to previous PuM techniques. This reduction leads to notable energy savings and performance improvements in bulk multiplications, while also providing flexibility to support up to 8-bit operand precision.

\textit{Simulation.} We developed an in-house simulator to model the performance and energy characteristics of Lama and all PuM baselines. The simulator leverages the HBM2 architectural configuration and the timing/energy parameters shown in Table~\ref{t:configuration} to calculate the number of commands, energy consumption, and latency, for bulk multiplications at 4-bit and 8-bit integer precision.

\textit{Baselines.} For comparison, we evaluate a baseline CPU on a real system equipped with an Intel\textsuperscript{\circledR} Xeon W-2245 CPU~\cite{intel}, utilizing AVX-512 Intel\textsuperscript{\circledR} Streaming SIMD Extensions for 8-bit multiplications. We also evaluate prior PuM approaches including pLUTo~\cite{pluto}, as a LUT-based technique, and SIMDRAM~\cite{simdram}, as a charge-sharing-based PuM technique. To ensure a fair comparison, the level of parallelism is consistent across all the baselines and our Lama implementation.

\subsection{Overheads}\label{overhead}
We assume an 8GB HBM with 4 layers (1GB per channel) as the memory configuration for Lama. Lama introduces two major components to the commodity DRAM: $(i)$ column counters and $(ii)$ mask logic. In addition, it includes a temporary buffer. These components were implemented using Verilog HDL and synthesized on the Synopsys Design Compiler with a 28nm technology library. In order to match the rate of column access time $t_{CCD} = 2$ns, the added bank-level components are configured to run at 500 MHz clock frequency. The area and power data obtained from the synthesis are scaled to 22nm to match the memory technology. We take into account the difference in manufacturing process between logic and DRAM similar to previous studies~\cite{technology_difference, transpim, drisa, fulcrum}, where the DRAM process incurs around 50\% additional area overhead for the logic.

Table~\ref{t:overhead} summarizes the area and power consumption of each unit. Each memory bank in Lama is equipped with sixteen column counters, one mask logic unit, and a temporary buffer. Assuming that all banks across the HBM2 channels are equipped with these units, the additional components consume approximately 1.32 mm\textsuperscript{2}, resulting in a 2.47\% area overhead. This is significantly lower than the area overheads reported in \cite{pluto, simdram}, thus preventing any significant loss in DRAM density and memory capacity.


\begin{table}[t!]
\caption{Summary of the area and power consumption of the added logic in the Lama architecture.}
\label{t:overhead}
\begin{center}
\resizebox{1.0\columnwidth}{!}{%
    \begin{tabular}{|c|cc|}
    \hline
    \textbf{Units}     &  \textbf{Area(um\textsuperscript{2} per Bank}) & \textbf{Power (mW) per Bank}        \\
    \hline
    Column Counter/Latch   &     5002.8                            &       1.49             \\
    Mask Logic             &     1628                              &       1.01             \\
    Temporary Buffer       &     3636.6                            &       3.76             \\
    Others                 &     19.73                             &       0.09             \\
    \hline
    \end{tabular}%
\quad
    \begin{tabular}{|c|c|}
    \hline
    \textbf{Lama}         &  \textbf{Area(mm\textsuperscript{2}})       \\
    \hline
    8GB HBM2              &     53.15                                   \\
    Overhead              &     1.32                                    \\
    \hline
    \end{tabular}%
}
\end{center}
\end{table}

\subsection{Evaluation}\label{eval_mult8b}
Previous PuM architectures have demonstrated promising results in executing bulk operations. However, these gains come at the cost of frequent ACT commands, which significantly increase energy consumption. Additionally, they face significant challenges in supporting operand sizes greater than 4-bit precision. For higher precision, their performance deteriorates due to the substantial increase in the number of issued commands, particularly for complex arithmetic operations.

To demonstrate Lama's higher performance in bulk multiplications, we compare it against existing proposals using 4-bit and 8-bit operand precisions. Table~\ref{t:cs1_eval_baselines} presents the latency, energy consumption, performance and command count for both precisions. To ensure a fair comparison, the level of parallelism is set to four for all schemes, which is consistent with the parallelism level used in the evaluation section of the pLUTo paper~\cite{pluto}. In Lama, this parallelism is achieved by executing operations across four different banks, where each bank performs operations required for one operand-coalesced batch, with one subarray per bank dedicated to LUT computation. Meanwhile, the other baselines employ subarray-level parallelism, using four subarrays within a single bank.

All techniques perform the same 1024 multiplication operations using 4 scalar operands, meaning that each bank (or subarray) handles computations related to one coalesced batch, with each bank or subarray executing 256 multiplication operations. Four key conclusions can be drawn from the results in Table~\ref{t:cs1_eval_baselines}.

First, Lama significantly reduces the number of ACT commands compared to other schemes. The ACT command count in Lama involves both reading the elements of $b$ from the source subarray and performing the LUT computation in the compute subarray. As precision increases from 4-bit to 8-bit, Lama requires the same ACT command count because row accesses are independent of operand precision. The only increase is in the number of read commands, which have a much lower energy impact compared to ACT commands.

Second, the SIMDRAM~\cite{simdram} baseline is based on the Triple Row Activation (TRA) mechanism in \cite{ambit}, which executes bitwise operations using a sequence of commands for MAJ/NOT operations. As shown in the study, each additional simultaneous row activation increases energy consumption by 22\% over a single-row activation. SIMDRAM also requires multiple execution cycles, and as operand precision increases, the number of cycles grows exponentially, leading to significant latency inefficiency. Lama achieves $13.7\times$ $(13.4\times)$ throughput for 4-bit (8-bit) precision and $5.8\times$ $(5.4\times)$ energy improvement over SIMDRAM.

Third, the pLUTo~\cite{pluto} baseline can perform 256 simultaneous bulk operations per subarray. However, multiple ACT commands are required to search all the LUT rows for the matching results, which increases energy consumption, particularly for 8-bit multiplication, where the number of rows activated rises exponentially. Lama shows $9.6\times$ $(8.3\times)$ energy savings over pLUTo in 4-bit (8-bit) multiplication. In terms of speedup, Lama achieves $3.8\times$ $(3.5\times)$ throughput for 4-bit (8-bit) precision. The decrease in throughput at 8-bit precision is due to the reduced parallelism and the extra cycles for filtering invalid data in the mask logic.

Fourth, compared to the CPU baseline for 8-bit multiplication, the LUT-based PuM architectures, Lama and pLUTo, show performance improvements of $3.8\times$ and $1.09\times$, respectively. For energy efficiency, all PuM architectures demonstrate greater energy savings than the CPU, with Lama obtaining the highest energy savings at $8\times$ over the CPU baseline.

Overall, Lama effectively handles multiplications of up to 8-bit operands while reducing the number of memory commands required to perform the computations in-memory, resulting in higher performance and a more energy-efficient solution for complex operations.

\begin{table}[t!]
\caption{Comparison of Lama for bulk multiplication vs. prior PuM works. All methods execute 1024 multiplications in both 4-bit and 8-bit integers with a parallelism level of 4. Results are calculated based on the HBM2 configuration provided in Table~\ref{t:configuration}.}
\label{t:cs1_eval_baselines}
\begin{center}
\resizebox{1.0\columnwidth}{!}{%
    \begin{tabular}{|c|c|c|c|c|}
    \hline
    \makecell{Methods}                        &  \textbf{pLUTo~\cite{pluto}}    &   \textbf{SIMDRAM~\cite{simdram}}   &    \textbf{Lama} &    \textbf{CPU}   \\
    \hline
    \multicolumn{4}{|c|}{{\textbf{INT-4 multiplication}}}  \\  
    \hline
    {Latency (ns)}              &              2240               &              7964                 &           583       &    -\\
    \hline
    {Energy (nJ)}               &              247.4             &              151.23                &           25.8      &    -\\
    \hline
    {Performance (GOPs/s)}      &              0.46               &              0.13                 &           1.75      &    -\\
    \hline
    {Num ACT commands}          &              1088               &              310                  &            8        &    -\\
    \hline
    {Num Total commands}        &              2176               &              465                  &           112       &    -\\
    \hline
    \multicolumn{4}{|c|}{{\textbf{INT-8 multiplication}}}  \\
    \hline
    {Latency (ns)}              &              8963              &              34065                &           2534      &    9760.4\\
    \hline
    {Energy (nJ)}               &              989.7             &              646.9               &           118.8     &    7900\\
    \hline
    {Performance (GOPs/s)}      &              0.11               &               0.03                &           0.4      &    0.1\\
    \hline
    {Num ACT commands}          &              4352               &               1326                &            8        &    -\\
    \hline
    {Num Total commands}        &              8704               &               1989                &           592      &    -\\
    \hline
    \end{tabular}%
}
\end{center}
\vskip -0.24in
\end{table}











\section{Case study 2: LamaAccel}
Attention-based models rely heavily on matrix multiplications to process input data, which involve numerous multiply and accumulate operations. These operations demand substantial data movement between memory and compute units. While LUT-based Processing-using-Memory (PuM) techniques can efficiently handle multiplication operations with negligible area overhead and minimal modifications to commodity memory, the accumulation process within memory presents a significant challenge. It requires extensive intra-bank (or subarray) data movement to reorganize data for accumulation, which diminishes the efficiency of PuM operations~\cite{transpim}. Many previous proposals address this challenge by integrating computational units within memory banks to perform accumulation and support hardware acceleration for DNNs~\cite{transpim,fulcrum,sal-pim}.

We propose LamaAccel, an HBM-based accelerator for large language models (LLMs) that efficiently addresses the challenge of handling accumulation in dot-product operations without requiring significant modifications to the memory organization. LamaAccel employs the \textbf{exponential quantization} technique based on the DNA-TEQ approach to simplify the dot-product of activations and weights into the sum of four terms (see Equation~\ref{eqn:conv_extended}). Each term consists in a base $b$ to the power of different exponents ($int_{A_{i}}, int_{W_{i}}$, and $int_{A_{i}} + int_{W_{i}}$), as explained in Section~\ref{Exponential_quant}. The weights and activations are encoded using the format $\{S_{A/W}, int_{A/W}\}$, where $S_{A/W}$ denotes the sign of the real value prior to quantization, and $int_{A/W}$ is a 3- to 7-bit integer. For the rest of this section, when we talk about the precision of a given layer, we are referring to the precision of the exponent. The dot-product operations can be efficiently implemented by counting the frequency of each exponent's occurrence, including the addition of exponents as seen in term 1. This approach transforms the problem of memory-intensive dot-product operations into more memory-friendly addition and counting tasks, which are efficiently handled by our Lama technique.


We leverage Lama to efficiently perform addition and counting operations directly within memory. After counting all exponents involved in computing an output activation, the result is transferred to the HBM logic die for post-processing, where the data is re-quantized and sent back as a new activation for the next layer. This execution flow offers a key advantage: the number of unique exponents to be counted per output neuron is much smaller than the total number of dot-product partial results that would typically be accumulated, especially in higher-precision layers. For instance, in a 6-bit precision layer, only $2^6$ unique exponents have to be counted, significantly reducing the number of data transfers to the logic die and, hence, reducing the energy overhead.

\subsection{LLM Layer Mapping}
LamaAccel further leverages HBM's pseudo-channels to pipeline the execution of encoder and decoder blocks in large language models (LLMs), enabling simultaneous parallel processing of multiple inferences. Each pseudo-channel is dedicated to executing an individual encoder or decoder block, with each layer following an input-stationary dataflow for efficient execution. For LLMs used in text generation tasks that consist of both encoder and decoder blocks, the throughput of decoder blocks is lower due to the serialized token execution. To maintain a balanced pipeline, LamaAccel allocates a greater number of resources (pseudo-channels) to the decoder blocks, thereby accelerating their execution during inference.

The layers within an encoder or decoder block are processed sequentially, utilizing the resources (i.e., banks) within each pseudo-channel. For fully-connected (FC) layers whose weights are statically known, the weights are pre-stored in the banks. For matrix-multiplication operations inside the self-attention blocks employing the $K$ and $V$ matrices, LamaAccel writes those matrices into the banks as if they were the weight parameters of FC layers, to compute the score matrix and the final attention output respectively.

Within each bank, mat-level parallelism is employed to process computations related to different output neurons with the same activation concurrently. The degree of parallelism ($p$) and thus the number of output neurons processed in parallel varies depending on the precision of each layer and is adjusted dynamically.

\subsection{Data Layout}\label{data_layout_casestudy2}
\begin{figure}[t!]
\centering
\includegraphics[width=0.99\columnwidth]{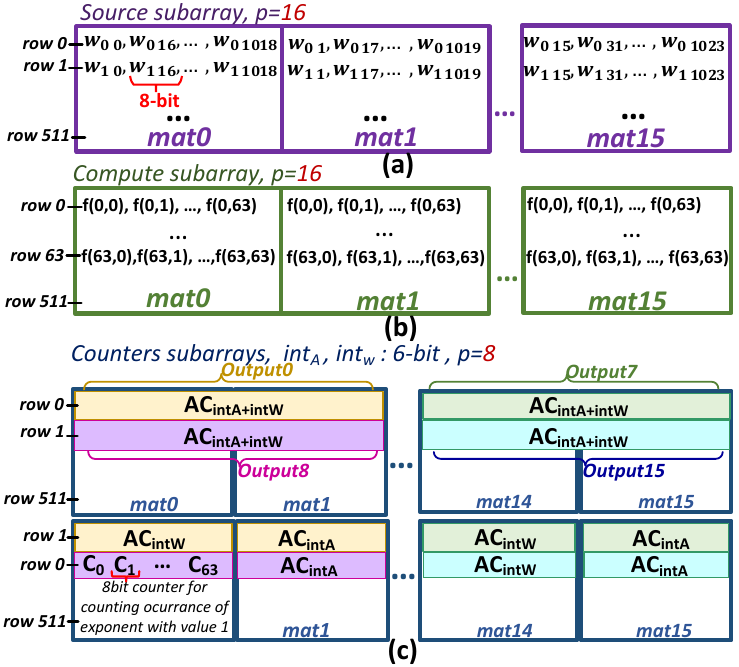}
\caption{LamaAccel data layout.}
\label{fig:LamaAccel_datalayout}
\vskip -0.15in
\end{figure}

LamaAccel employs the subarrays of a DRAM bank for three distinct purposes during the execution of a DNN layer. First, a group of subarrays (\textit{source subarrays}) is dedicated to storing the exponentially quantized weights of a layer. The exponent for each weight is pre-computed and stored statically. Each weight is stored in an 8-bit format as $(S_{W}, int_{W})$, with padding in the exponent. This fixed 8-bit format ensures compatibility with all possible precision levels across layers. Figure~\ref{fig:LamaAccel_datalayout}a (shown in purple) illustrates the organization of weights in one source subarray. Since each layer's execution follows an input-stationary dataflow, all the weights corresponding to an input index are stored in the same row of the subarray. The weights in each row are arranged so that in every column access, 16 weights for 16 different output neurons are fetched from all mats. For example, when activating row 0, corresponding to input activation 0, the first column access retrieves weights $W_{0,0}$ to $W_{0,15}$ corresponding to the first 16 output neurons.

The second group of subarrays (\textit{compute subarrays} shown in Figure~\ref{fig:LamaAccel_datalayout}b in green) is dedicated to storing the LUT data required during the computation of the sum of exponents ($int_{A} + int_{W}$) corresponding to the first term in Equation~\ref{eqn:conv_extended}. All exponent sums are pre-computed and stored as 8-bit padded results. For layers with up to 6-bit exponents, each mat can accommodate one LUT. In each row of a mat, all possible cases for a given input activation exponent ($int_{A}$) are covered, ensuring the maximum degree of parallelism ($p=16$). However, for layers with 7-bit precision exponents, storing all possible values related to a single input activation in the same row requires two mats, reducing the degree of parallelism to $p=8$. This is because a 7-bit exponent can yield $2^7$ possible values for the sum of exponents, which must be stored as 8-bit results. Consequently, a row needs 1024 bits to store these values, requiring two mats to accommodate the data.

The third group of subarrays (\textit{counters subarrays} shown in Figure~\ref{fig:LamaAccel_datalayout}c in blue) is dedicated to tracking the occurrence of exponents for output neurons during the counting process. Each subarray updates the occurrences for $p$ output neurons simultaneously. Each output neuron ($ON_{i}$) requires three separate arrays of counters ($AC_{i}$) to track occurrences of the exponents ($int_{A_{i}}$, $int_{W_{i}}$, and $int_{A_{i}} + int_{W_{i}}$). The number of counters in each array varies depending on the layer's precision. For example, for 4-bit exponents, the array counter for $int_{A}$ contains $2^4$ counters. To determine the appropriate bitwidth for the counters, we analyzed the maximum occurrences of all possible exponents across the layers of our evaluated DNNs. This analysis showed that an 8-bit size per counter is sufficient to track the counts for each term without encountering numerical instability.



Each row of the counters subarray contains the array counters for different output neurons, allowing their occurrences to be updated simultaneously. Figure~\ref{fig:LamaAccel_datalayout}c illustrates the organization of the array counters in a subarray. For a layer precision of 3 and 4 bits, all array counters for an output neuron fit within a single mat, maintaining the parallelism degree of $p=16$, meaning that occurrences for the same term across 16 output neurons can be updated simultaneously. However, with 5-bit precision, two subarrays are required to fit all array counters while still preserving $p=16$. In the case of 6-bit precision, the array counters are distributed across two subarrays, providing a parallelism degree of $p=8$. For the highest precision of 7-bit, the parallelism degree decreases further to $p=4$, with the array counter placement following a similar organization to the 6-bit precision case. The total number of subarrays required for a given layer is determined by the number of output neurons and the parallelism level $p$.

\subsection{Execution Flow}\label{ExecutionFlow_cs2}
This section details how LamaAccel computes each term in Equation~\ref{eqn:conv_extended}. The bank structure in LamaAccel is similar to that in \textit{Case Study 1}, with two key enhancements. First, to accommodate the encoded input activation value {$S_{A_{i}}, int_{A_{i}}$} currently being processed for all output neurons, each bank is equipped with an 8-bit activation buffer. Second, each column address counter/latch is enhanced with an XNOR gate and a de-multiplexer, allowing it to increment or decrement the number of occurrences for each exponent based on the XNOR result of the input and weight signs. Additionally, the latch size of the column counters is extended from 5 bits (as used in commodity memories) to 8 bits, allowing it to store the number of occurrences of a given exponent. This extra size of the latch for each counter has been accounted for in the area overhead calculation of the column address counters, as discussed in Section~\ref{overhead}.


LamaAccel’s dataflow follows an \textit{input-stationary} approach, where the input activation remains stationary across the processing elements while computations for different output neurons are performed. In this scheme, for each iteration, a given input activation is broadcast to all banks within a pseudo-channel and stored in the activation latch, with each bank responsible for processing a subset of output neurons. By keeping the input activation constant and distributing the weight-related computations across banks, LamaAccel minimizes the need for data transfers and avoids the replication of weights across banks. This approach contrasts with weight-stationary dataflows, which replicate weights across multiple banks, leading to higher memory overhead. The memory controller in the logic die orchestrates the commands for output neuron computations in each bank based on the current input activation, ensuring efficient parallel execution across all banks in the pseudo-channel.


The execution flow is consistent across terms 1, 2 and 3, so we focus on detailing the process for the first term, which also involves the addition of exponents. The flow for terms 2 and 3 is similar but does not include the addition of exponents. The execution process in LamaAccel is divided into three key steps: (1) acquiring the weights associated with the output neurons being processed, (2) computing the sum of exponents, and (3) updating the number of occurrences of exponents. Each step operates with a different degree of parallelism ($p$), depending on the number of simultaneous operations that can be performed. This flexibility allows for efficient utilization of resources and optimized performance based on the specific requirements of each step.

\textbf{Step 1: Weight Acquisition.} In this stage, the memory controller issues an ACT command to activate the row $\#i$th in the \textit{source subarray}, based on the \textit{positional index} $i$ of the currently processed activation exponent $int_{A_{i}}$. This row contains 1024 encoded weights ($W_{i,0}$ to $W_{i,1023}$) in HBM2, corresponding to 1024 different output neurons for the current input activation. Following the ACT command, an internal column access (ICA), is executed to fetch 16 weights, which are then stored in the temporary buffer. This stage mirrors the execution steps (\circled{1} and \circled{2}) for bulk multiplication, illustrated in Figure~\ref{fig:case_study1}, and supports a degree of $p=16$, facilitating efficient weight retrieval. Once the initial 16 weights are fetched, the corresponding computations for these weights are performed. Subsequent weights can be fetched without issuing a new ACT command, as the row remains open. This allows for retrieving 16 weights with only a single ICA command each time, reducing the overhead and enhancing energy efficiency.

\textbf{Step 2: Computing the Sum of Exponents.} The process begins by performing a LUT activation for the row $\#int_{A_{i}}$th in the compute subarray, based on the \textit{value} of the activation exponent ($int_{A_{i}}$). Following this, the column address counters initiate the LUT retrieval operation using $p$ weight exponents ($int_{W}$) stored in the temporary buffer. Each mat outputs the sum of exponents, which is then stored in the temporary buffer. This continues until the sum for all weight exponents fetched in the first stage is fully computed. For layers with precision lower than 7-bit, LamaAccel supports the full parallelism degree ($p=16$), in which the mask logic is bypassed and all retrieved values are fetched during one ICA. However, in the case of 7-bit precision, $p$ is equal to 8 and so eight weight exponents are broadcast to all column address counters/latches -with each pair of consecutive column counters processing the same weight exponent- while the mask logic filters out invalid results. Overall, two ICAs are required to compute the sums for all stored weight exponents in the temporary buffer for 7-bit precision, and only one ICA for lower precisions. As previously mentioned, each iteration handles computations for a specific input activation across all output neurons in the layer. Since each row in the compute subarray contains all possible sum results for the current value of the activation exponent, the same row remains open during these iteration. This eliminates the need for repeated ACT commands when processing new sets of output neurons, requiring only ICA commands, which improves energy efficiency.

\begin{figure*}[t!]
\centering
\includegraphics[width=1.0\textwidth]{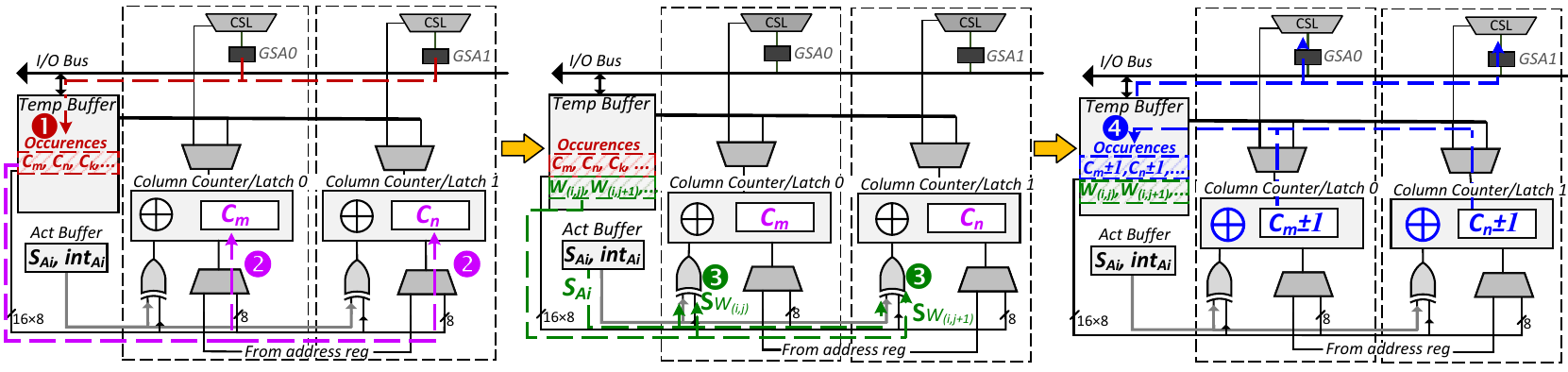}
\caption{Counting the occurrence of exponents.}
\label{fig:step3}
\vskip -0.15in
\end{figure*}

\textbf{Step 3: Counting the occurrence of Exponents.} In this step, the number of exponent occurrences is updated based on the XNOR result of the signs of the weights and input activations, denoted as $S_{W_{i}}S_{A_{i}}$. If the XNOR result is 1, the sign multiplication is negative and the corresponding occurrence count is decremented; otherwise, it is incremented. As shown in Figure~\ref{fig:step3}, the process is as follows: \textit{First}, the memory controller issues the ACT command to activate the row in the counters subarray where the array counter ($AC_{A+W}$) for 16 output neurons is stored. The precision of each layer determines whether the selected array counters fit within one row ($p=16$) or span multiple rows ($p=8,4$). If $p<16$ the counting step is repeated to cover all 16 output neurons processed in the previous steps. The corresponding occurrences are fetched and stored into the temporary buffer through one ICA (\circled{1}). \textit{Second}, the fetched occurrences are loaded into the column address counter/latch for being updated in the next cycle, as shown in \circled{2}. \textit{Third}, the XNOR of $S_{W_{i}} and S_{A_{i}}$ is computed for each occurrence by retrieving the signs from the temporary and activation buffers. Then, the counters are updated based on the XNOR result (\circled{3}). \textit{Forth}, the updated occurrences are written back into the temporary buffer. Thanks to the open-page policy, a new ACT command is not required, as the row is kept open (\circled{4}).

As described in previous sections, during a single ICA, each mat in the subarray provides an 8-bit value, selected through the column address counters/latches during the column selection process. During this stage, For layers where $p<16$, not all fetched occurrences from the 16 mats align with the target exponent occurrence counter in the array counter. This mismatch occurs because, depending on the layer's precision, the array counter for a given output neuron may span multiple mats. To address this, LamaAccel ensures that only column address counters with valid fetched occurrences perform the count up/down operation, while counters associated with non-valid occurrences remain in latch mode, preserving the fetched values without modification.

After completing the computations for term 1 of the equation, LamaAccel proceeds to compute terms 2 and 3. If the array counters for these terms reside in the same row as the array counter $A+W$, no new ACT command is required to open a new row in the counters subarray. Otherwise, new ACT commands are issued to access the corresponding array counters for these terms.

Once computations for all three terms for a given set of output neurons are complete, a PRECHARGE command closes the opened row in the counters subarray. This step is crucial for enabling the system to move on to the next set of output neurons, as each row in the counters subarray corresponds to a specific set of neurons. These steps are repeated within the current input activation iteration until all computations for all output neurons are completed. After processing all input activations for a layer, the values in the array counters are transferred to the logic die for post-processing, as detailed in Equation~\ref{eqn:conv_extended}, to produce the input activations for the next layer. Importantly, the next layer's computations can begin in a separate pseudo-channel containing the next layer's weights as soon as the first input activation is ready. This overlapping of post-processing and the next layer's computation helps reduce overall latency.

\begin{table}[t!]
\caption{Baseline accuracy vs accuracy after performing exponential quantization for the evaluated LLM models. The average bitwidth is the mean each layer's exponents.}
\label{t:LLMs}
\vskip -0.20in
\begin{center}
\resizebox{1.0\columnwidth}{!}{%
    \begin{tabular}{|c|c|c|c|c|c|}
    \hline
    \textbf{Network}       & \textbf{Task}        & \textbf{Baseline Acc}                  & \textbf{Quantized Acc} & \textbf{Avg bit} & \textbf{max SL}          \\
    \hline
      \multirow{2}{*}{\textbf{BERT-Base}}  &   SQuAD1    &  88.68\% (F1)                  & 88.13\%            &              6.45     & 384                      \\
    
        &   GLUE-SST2                      &  91.70\% (Exact match)                       & 90.82\%            &              3.48     & 128                      \\
    \hline
    \multirow{2}{*}{\textbf{BART-Large}}   &   CNN-DM    &  29.98\% (F1 Rouge L)          & 29.13\%            &              5.71     & 142                      \\
    
       &   MNLI                            &  90.17\%  (F1)                               & 89.34 \%           &              4.88     & 1024                     \\
    \hline
    \textbf{GPT-2-Small}  &   IMDB         &  94.46\%  (F1)                               & 94.16\%            &              6.03     & 1024                     \\
    \hline
    \end{tabular}%
}
\end{center}
\vskip -0.20in
\end{table}

\subsection{Methodology}
The hardware characteristics for LamaAccel are the same as those for the Lama evaluation platform, summarized in Table~\ref{t:configuration}. For LamaAccel we evaluate three widely-used attention models: BERT\cite{bert}, BART\cite{bart}, and GPT-2\cite{gpt-2}, across a range of representative NLP tasks, including text classification (IMDB~\cite{imdb} and MNLI~\cite{mnli}), question answering (SQuAD1.1~\cite{squad}), text summarization (CNN-DM~\cite{cnn-dm,cnn-dm2}), and sentimental analysis (GLUE-SST2~\cite{glue-sst2}). The BERT model corresponds to the base model consisting of 12 encoder blocks, while the BART model is the large version with both encoder and decoder blocks (12 encoders and 12 decoders). For GPT-2, we use the GPT-2 small model, which consists of 12 decoder blocks. All workloads are implemented in PyTorch using the HuggingFace Transformers library~\cite{huggingface}. 

The models are exponentially quantized, with quantization parameters and layer precision determined using the search algorithm described in \cite{DNA-TEQ}, based on Equation~\ref{eqn:conv_extended}. The quantization ensures less than 1\% accuracy loss compared to the baseline models without requiring fine-tuning. Table~\ref{t:LLMs} presents the baseline accuracy in FP32 and the accuracy after quantization. Additionally, for each model, various tasks are evaluated, and the average exponent bitwidth for each task is also reported. 

For performance and energy evaluations, we assume the workloads operate at their maximum sequence length, as indicated in Table~\ref{t:LLMs} under "Max SL". We employ the Lama simulator configured as described in the methodology of Case Study 1. The area overhead is consistent with that reported in Section~\ref{overhead}, with an additional area of $0.01 \ \text{mm}^2$ added to the HBM2 due to the extra components described in Section~\ref{ExecutionFlow_cs2}.


For comparison with GPUs, we use as baseline a Nvidia RTX A6000. To measure GPU performance, we focus on the kernel execution time, excluding the data initialization overhead. Energy consumption is measured using Python bindings for {\fontfamily{cmss}\selectfont nvml} API~\cite{NVIDIA}. 

For comparison with a TPU, we extend the ScaleSim~\cite{scale} simulator to model a TPU-like architecture using the specifications from the Google Edge TPU Coral~\cite{tpu-edge}, which has a $64 \times 64$ systolic array, a chip area of $43.29 \ \text{mm}^2$, and a frequency of 480 MHz, similar to the added components in LamaAccel's banks. We assume an 8MB global on-chip SRAM~\cite{tpu-edge-sram} and 1GB of off-chip LPDDR4 memory~\cite{tpu-edge-ddr4}, with all layers set to 8-bit integers.

For comparison with previous PuM accelerators we implement pLUTo~\cite{pluto}, described in Section~\ref{pluto_background}, using the same dataflow and layer mapping as LamaAccel, with subarray-level parallelism set to 16 to match LamaAccel's bank-level parallelism. Given that pLUTo supports only 4-bit precision, we assume that all layers in the LLM workloads are uniformly quantized to 4-bit precision for a fair comparison. Although this uniform quantization does not guarantee that accuracy drops will remain below 1\% in the evaluated workloads, we overlook this limitation to maintain consistency with pLUTo's constraints.

\subsection{Evaluation}
\begin{figure}
    \centering
    \begin{subfigure}{\columnwidth}
        \centering
        \includegraphics[height=2.8cm, width=8.1cm]{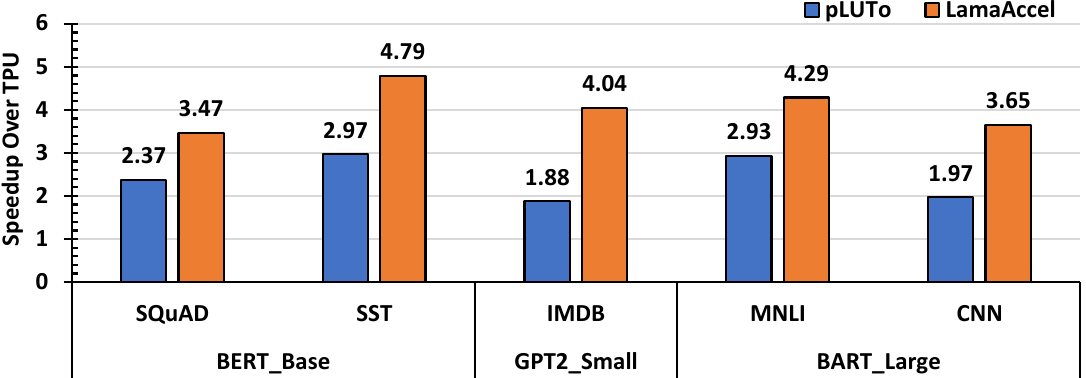}
    \end{subfigure}
    \begin{subfigure}{\columnwidth}
        \centering
        \includegraphics[height=2.9cm, width=8.2cm]{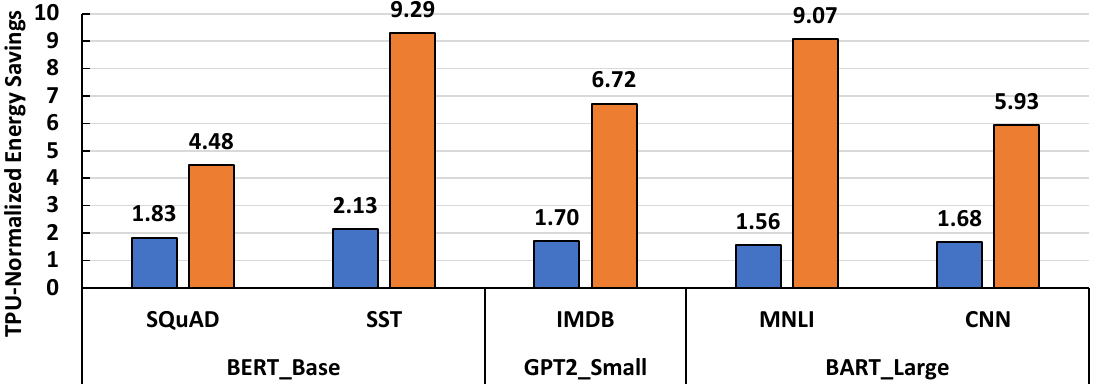}
        \label{fig:third}
    \end{subfigure}
    \caption{LamaAccel and pLUTo speedup and energy savings, normalized to TPU.}
    \vskip -0.20in
    \label{fig:case_study2_evaluation_tpu}
\end{figure}

Figure~\ref{fig:case_study2_evaluation_tpu} illustrates the speedup and normalized energy savings of a PuM baseline (pLUTo) and LamaAccel over TPU for three LLM workloads across different NLP tasks. LamaAccel consistently achieves significant speedups, from $3.4\times$ (BERT for SQuAD1) to $4.7\times$ (BERT for SST2), with an average improvement of $4.1\times$. Compared to pLUTo, LamaAccel delivers an average speedup of $1.7\times$ across all workloads. Notably, since pLUTo only supports up to 4-bit operand precision, all workloads in pLUTo are executed at 4-bit precision. In contrast, LamaAccel processes most workloads at higher average precision, yet still outperforms pLUTo in all cases, demonstrating better speedups even with higher precision.

For tasks performed with lower average bitwidth, the speedup achieved is generally higher. In all tasks except for the BART CNN text summarization, each pseudo-channel is assigned to a single encoder or decoder block. In BERT for SST2, with the lowest average bitwidth of 3.4, the speedup is higher than the rest of the workloads. For BART's text summarization on the CNN dataset, decoder blocks generate tokens sequentially, one token at a time, making the decoder blocks a bottleneck during inference. The bottleneck arises because each step depends on generating the previous token, which slows down the process. To address this, we allocate 2 pseudo-channels for the encoder blocks and the remaining resources for faster execution of the decoder blocks, ultimately achieving a speedup of $3.6\times$.

Energy savings in LamaAccel are closely linked to the average precision of each LLM. As the precision of the model layers becomes lower, LamaAccel shows greater energy savings compared to the TPU baseline. BERT on the GLUE-SST2 task, which has the lowest average precision of 3.48 (see Table~\ref{t:LLMs}), achieves the highest energy saving of $9.2\times$. Conversely, BERT on the SQuAD task, with the highest average precision, results in a lower energy saving of $4.4\times$ compared to TPU. LamaAccel reduces the higher precision overhead by minimizing the number of ACT commands and exploiting the open-page policy, allowing for simultaneous operations across multiple output neurons. In the weight acquisition and sum computation stages, it reuses the already activated row, requiring only additional read commands. Since the energy cost of column accesses is much lower than that of new ACT commands, LamaAccel exhibits better energy efficiency than pLUTo. Additionally, during the counting phase, the mapping of array counters is designed to ensure that even with increased precision, the system maintains maximum efficiency in processing output neurons.

\begin{figure}[t!]
\centering
\includegraphics[height=3.5cm, width=8.5cm]{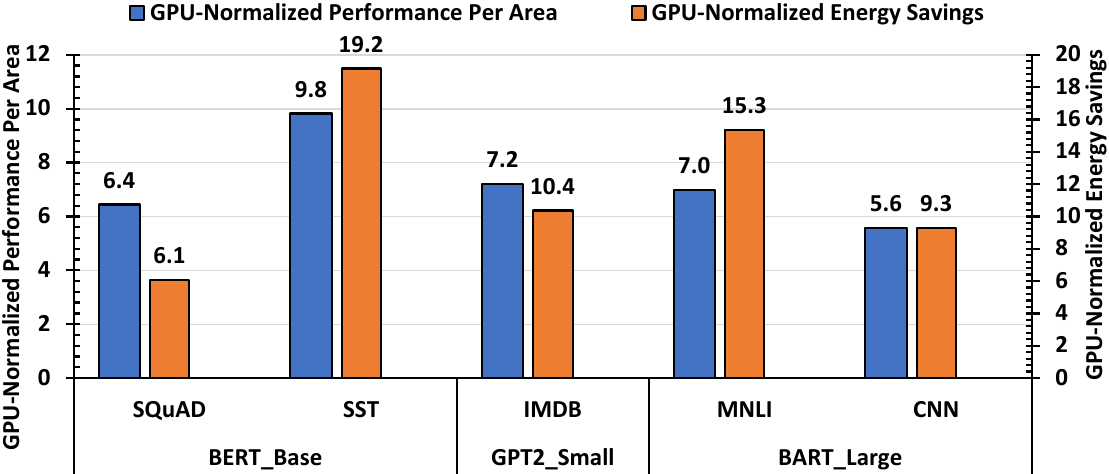}
\caption{Speedup and energy savings of LamaAccel normalized to the GPU. In the case of energy savings, higher is better.}
\label{fig:case_study2_evaluation_gpu}
\vskip -0.15in
\end{figure}

In Figure~\ref{fig:case_study2_evaluation_gpu} LamaAccel performance per area is shown relative to the GPU baseline. The NVIDIA RTX A6000 features a die size of $628 \ \text{mm}^2$ on an $8 \ \text{nm}$ process, whereas LamaAccel occupies $53.15 \ \text{mm}^2$ on a $21 \ \text{nm}$ node, equivalent to the area of a single HBM2 stack. For this comparison, LamaAccel's technology node is not scaled to the GPUs, inherently favoring the GPU. Despite this, LamaAccel achieves an average of $7.2\times$ higher performance per area across workloads. In energy efficiency, LamaAccel outperforms the GPU baseline by $6.1\times$ to $19.2\times$, with higher savings in tasks using lower precision. While LamaAccel's inference throughput is lower than that of the GPU, primarily due to the limited resources in a single HBM2 stack compared to a high-end GPU, its scalable architecture allows throughput to scale linearly with additional HBM2 stacks, bringing it closer to the performance of high-end GPUs.

\section{Conclusions}
In this paper, we proposed Lama, a lightweight adaptive memory access mechanism for executing bulk arithmetic operations in-memory for SIMD workloads. Lama is a LUT-based Processing-in-Memory (PuM) approach that enables parallel, column-independent accesses within each mat of a DRAM subarray. It supports operand precision of up to 8 bits while optimizing latency and energy efficiency by leveraging intrinsic DRAM features like mat-level parallelism and the open-page policy. Lama integrates seamlessly into existing DRAM architectures, incurring only a minimal area overhead of 2.47\%. In addition, Lama reduces the number of required commands for bulk operations by fivefold compared to state-of-the-art techniques like pLUTo. 

We also introduced LamaAccel, an HBM-based accelerator designed for large language models (LLMs) that provides efficient execution without requiring modifications to the DRAM timing parameters. LamaAccel tackles the challenge of accumulation in dot-product operations within memory by utilizing exponential quantization to simplify computations of activations and weights. Evaluations across various workloads demonstrate that LamaAccel significantly outperforms alternative platforms, including TPU, GPU, and pLUTo, in terms of both performance and energy efficiency.

\section{Acknowledgement}\label{acknowledgement}
This work has been supported by the CoCoUnit ERC Advanced Grant of the EU’s Horizon 2020 program (grant No 833057), the Spanish State Research Agency (MCIN/AEI) under grant PID2020-113172RB-I00, the Catalan Agency for University and Research (AGAUR) under grant 2021SGR00383, and the ICREA Academia program.

\bibliographystyle{IEEEtranS}
\bibliography{references}

\end{document}